\documentclass[%
aps,
prd,
amssymb,
amsmath,
amsfonts,
eqsecnum,
showpacs,
nofootinbib,
floatfix,
twocolumn,
twoside,
a4paper,
superscriptaddress,
longbibliography
]{revtex4-1}

\usepackage[T3,T1]{fontenc}
\usepackage{mathrsfs} 
\usepackage{bm} 
\makeatletter
\@ifclassloaded{beamer}
  { 
    \typeout{UsePackages: Detected beamer}
    \usepackage{tgheros}
    
  }
  { 
    \typeout{UsePackages: Did not detect beamer}
    \ifx\asybeamer\undefined 
    \typeout{UsePackages: Detected article}
    \usepackage[varg]{txfonts} 
    \usepackage{tgtermes} 
    \else 
    \typeout{Fonts: Detected asy for beamer}
    \usepackage{tgheros}
    
    \fi
  }
\usepackage{microtype} 
\def\MT@register@subst@font{
  \MT@exp@one@n\MT@in@clist\font@name\MT@font@list
  \ifMT@inlist@\else\xdef\MT@font@list{\MT@font@list\font@name,}\fi}
\makeatother

\usepackage{graphicx} %
\usepackage[x11names,svgnames,rgb]{xcolor} %
\usepackage{xspace}
\usepackage{braket}
\usepackage{accents}
\usepackage{siunitx}
\usepackage{mathtools}
\DeclareSymbolFontAlphabet{\mathrm}{operators}

\definecolor{CiteColor}{rgb}{0.18039, 0.18824, 0.57255}
\definecolor{UrlColor} {rgb}{0.741, 0.173, 0.000}
\definecolor{DarkUrlColor} {rgb}{0.500, 0.110, 0.000}
\definecolor{LinkColor}{rgb}{0.25098, 0.47843, 0.04706}

\makeatletter %
\newcommand{\ShowFont}{%
  \typeout{The main font is \f@encoding \space \f@family \space %
    \f@series \space \f@shape \space at \f@size pt.}%
  \typeout{The math font sizes are \tf@size pt (main), \sf@size pt %
    (script), and \ssf@size pt (scriptscript).}%
  \typeout{The linewidth is \the\linewidth}} %
\makeatother %

\usepackage{xargs}

\usepackage{graphicx}
\usepackage{dcolumn}
\usepackage{bm}
\usepackage{tabularx}
\usepackage{multirow}
\usepackage{capt-of}
\usepackage{color}
\usepackage{url}
\usepackage[utf8]{inputenc}
\usepackage{placeins}

\usepackage[nolist,nohyperlinks]{acronym}
\usepackage{xspace}
\usepackage[colorlinks]{hyperref}

\usepackage[caption=false]{subfig}

\DeclareMathAlphabet{\mathbfsf}{\encodingdefault}{\sfdefault}{bx}{sl}

\newcommand{\ellmax}{\ell_{\text{max}}}

\newcommand{\be}{\begin{equation}}
\newcommand{\ee}{\end{equation}}
\newcommand{\bea}{\begin{eqnarray}}
\newcommand{\eea}{\end{eqnarray}}

\newcommand{\phIMR}{\textsc{IMRPhenom}\xspace}

\newcommand{\ppvtwo}{\textsc{IMRPhenomPv2}\xspace}

\newcommand{\phenX}{\textsc{IMRPhenomX}\xspace}
\newcommand{\phXF}{\textsc{IMRPhenomX}\xspace}

\newcommand{\phPvtwo}{\textsc{IMRPhenomPv2}\xspace}

\newcommand{\phPvthreehm}{\textsc{IMRPhenomPv3HM}\xspace}
\newcommand{\phXHM}{\textsc{IMRPhenomXHM}\xspace}

\newcommand{\phXPHM}{\textsc{IMRPhenomXPHM}\xspace}
\newcommand{\phT}{\textsc{IMRPhenomT}\xspace}
\newcommand{\phTHM}{\textsc{IMRPhenomTHM}\xspace}

\newcommand{\phTPHM}{\textsc{IMRPhenomTPHM}\xspace}

\newcommand{\NRSur}{\textsc{NRSur7dq4}\xspace}

\newcommand{\seobnrvforphm}{\textsc{SEOBNRv4PHM}\xspace}
\newcommand{\ph}{\textsc{IMRPhenom}\xspace}

\newcommand{\Msun}{M_\odot}
\newcommand{\chieff}{\chi_\mathrm{eff}}
\newcommand{\chip}{\chi_\mathrm{p}}
\newcommand{\maxL}{\max\mathcal{L}}

\definecolor{dodgerblue}{HTML}{1E90FF}
\definecolor{viennared}{HTML}{DA0A14}
\definecolor{ctorange}{HTML}{FF6C0C}
\definecolor{granadagreen}{HTML}{078931}
\definecolor{wales}{HTML}{ff0038}
\definecolor{valenciacfred}{HTML}{ee3524}
\definecolor{barcelonafcgold}{HTML}{edbb00}
\definecolor{jam}{HTML}{A50B5E}
\definecolor{austriawien}{HTML}{441678}

\AtBeginDocument{%
  \hypersetup{
    citecolor=dodgerblue,
    linkcolor=dodgerblue,   
    urlcolor=dodgerblue}}

\newcommand{\UIB}{Departament de F\'isica, Universitat de les Illes Balears, IAC3 -- IEEC, Crta. Valldemossa km 7.5, E-07122 Palma, Spain}

\newcommand{\AEI}{Max Plank Institut für Gravitationsphysik (Albert Einstein Institut), Am M\"uhlenberg 1, Potsdam, Germany}

\newcommand{\mymacro}[1]{{\ensuremath{#1}}}
\newcommand{\mOne}{\mymacro{85^{+21}_{-14}}}      
\newcommand{\mTwo}{\mymacro{66^{+17}_{-18}}}   
\newcommand{\mTotal}{\mymacro{150^{+29}_{-17}}}    
\newcommand{\mChirp}{\mymacro{64^{+13}_{-8}}}    
\newcommand{\massRatio}{\mymacro{0.79^{+0.19}_{-0.29}}}    
\newcommand{\mFinal}{\mymacro{142^{+28}_{-16}}}    
\newcommand{\aFinal}{\mymacro{0.72^{ +0.09}_{-0.12}}}    
\newcommand{\mTotalMinusMfinal}{\mymacro{7.6^{+2.2}_{-1.9}}}    
\newcommand{\aOne}{\mymacro{0.69^{ +0.27}_{-0.62}}}       
\newcommand{\aTwo}{\mymacro{0.73^{ +0.24}_{-0.64}}}       
\newcommand{\thOne}{\mymacro{81^{+64}_{-53}}}             
\newcommand{\thTwo}{\mymacro{85^{+57}_{-55}}}             
\newcommand{\chiEff}{\mymacro{0.08^{+0.27}_{-0.36}}}      
\newcommand{\chiP}{\mymacro{0.68^{+0.25}_{-0.37}}}        

\allowdisplaybreaks[1]

\begin{document}

\title[GW190521]
{A detailed analysis of GW190521 with phenomenological waveform models}


\author{H\'{e}ctor Estell\'{e}s}
\affiliation{\UIB}

\author{Sascha Husa}
\affiliation{\UIB}

\author{Marta Colleoni}
\affiliation{\UIB}

\author{Maite Mateu-Lucena}
\affiliation{\UIB}

\author{Maria de Lluc Planas}
\affiliation{\UIB}

\author{Cecilio Garc{\'i}a-Quir{\'o}s}
\affiliation{\UIB}

\author{David Keitel}
\affiliation{\UIB}

\author{Antoni Ramos-Buades}
\affiliation{\AEI}
\affiliation{\UIB}

\author{Ajit Kumar Mehta}
\affiliation{\AEI}

\author{Alessandra Buonanno}
\affiliation{\AEI}

\author{Serguei Ossokine}
\affiliation{\AEI}

\date{\today}

\begin{abstract}
In this paper we present an extensive analysis of the GW190521 gravitational wave event with the current (fourth) generation of phenomenological waveform models for binary black hole coalescences. GW190521 stands out from other events since only a few wave cycles are observable. This leads to a number of challenges, one being that such short signals are prone to not resolve approximate waveform degeneracies, which may result in multi-modal posterior distributions. The family of waveform models we use includes a new fast time-domain model (\phTPHM), which allows us extensive tests of different priors and robustness with respect to variations in the waveform model, including the content of spherical harmonic modes. We clarify some issues raised in a recent paper \cite{Nitz:2020mga}, associated with possible support for a high-mass ratio source, but confirm their finding of a multi-modal posterior distribution,
albeit with important differences in the statistical significance of the peaks. In particular, we find that the support for both masses being outside the PISN mass-gap, and the support for an intermediate mass ratio binary are drastically reduced with respect to what \cite{Nitz:2020mga} found.
We also provide updated probabilities for associating GW190521 to the potential electromagnetic counterpart from ZTF~\cite{Graham:2020gwr}.
\end{abstract}


\maketitle

\section{Introduction}
\label{sec:Introduction}

GW190521 \cite{Abbott:2020tfl,Abbott:2020mjq} is a uniquely stimulating gravitational wave (GW) event: it challenges our understanding of astrophysical formation channels of black holes, the accuracy of our waveform models, and our methods for data analysis. The signal found is a very short transient with a duration of only approximately 0.1 s, and around four cycles in the frequency band 30--80\,Hz \cite{Abbott:2020tfl}. The source of the signal was originally identified as most likely being the merger of a binary black hole (BBH) system with a total mass of about 150 solar masses in the source frame, the highest-mass merger observed to date. Furthermore, at least the more massive component was identified as having a very high probability of being inside the pair instability supernova (PISN) mass gap \cite{Woosley:2016hmi}. In addition, a potential electromagnetic counterpart has been identified \cite{Graham:2020gwr}, although its association with the GW event is not considered robust \cite{Graham:2020gwr,Ashton:2020kyr,Palmese:2021wcv}.

For such short signals it is however not surprising if GW waveforms corresponding to different source parameters fit the observed data equally well, and indeed already the original publication \cite{Abbott:2020mjq} by the LIGO Scientific and Virgo collaborations (LVC) discussed a wide range of possible alternative sources, and recent papers have proposed interpretations including that of a highly eccentric collision \cite{CalderonBustillo:2020odh,gayathri2020gw190521,Romero_Shaw_2020}, a Boson star merger \cite{CalderonBustillo:2020srq},
a high-mass black hole--disk system \cite{Shibata:2021sau},
or the first instance of an intermediate mass-ratio inspiral \cite{Nitz:2020mga}.
The latter paper found a tri-modal posterior distribution, whose modes required a careful choice of priors and sampler settings to be resolved when running with the precessing frequency-domain model \phXPHM \cite{Pratten:2020ceb}, which was developed recently, involving some of us.

For high mass ratios, waveform models are not yet calibrated to numerical relativity (NR) simulations of precessing systems, and even the coverage offered by aligned-spin NR waveforms is sparse when compared to approximately equal masses. Therefore, modelling and extrapolation effects are expected to be significant and the impact of waveform systematics in this region of parameter space is still poorly understood.
Indeed the possibility to choose among different precession prescriptions in \phXPHM represents a useful tool to investigate the impact of different modelling approximations on parameter estimation. Frequency-domain waveform models such as \phXPHM and its predecessors \ppvtwo \cite{Hannam:2013oca, Bohe:PPv2} and \phPvthreehm \cite{Khan:2019kot} use a number of common approximations (see \cite{Ramos-Buades:2020noq} for a recent discussion), in particular the ``twisting-up'' method to represent precession effects starting from NR-calibrated aligned-spin waveforms \cite{Hannam:2013oca, Bohe:PPv2}, and the stationary phase approximation (SPA). Both strategies allow to significantly accelerate waveform evaluation. The SPA is formally valid only in the slowly-evolving inspiral phase, and its continuation into the highly-dynamical merger-ringdown regime leads to inaccuracies that are likely to be particularly relevant for short-lived signals where only a few cycles around merger-ringdown are observed. 

In order to gain a better understanding of the impact of different modelling approximations on parameter estimation results, we reanalyze GW190521 with different variants of \phXPHM \cite{Garcia-Quiros:2020qlt,Garcia-Quiros:2020qpx,Pratten:2020ceb} and the new phenomenological time-domain model \phTPHM \cite{Estelles:2020osj, Estelles:2020twz,estelles2021new}. Unlike its frequency-domain counterpart, \phTPHM does not resort to the SPA approximation and offers a number of improvements in the description of precession effects both in the inspiral and merger-ringdown regimes.
In this paper we will systematically compare results obtained with both models, following the strategy of a closely related paper \cite{mateulucena2021adding} presenting a complete re-analysis of the GWTC-1 catalog \cite{LIGOScientific:2018mvr}, as well as of another publication specializing on GW190412 \cite{LIGOScientific:2020stg,Colleoni:2020tgc}, where similar systematic comparisons were carried out.

The purpose of this paper is twofold. First, we will clarify some of the challenges encountered in the analysis of high-mass non-vanilla GW events such as GW190521, in particular in terms of waveform systematics and robust Bayesian sampling. To this end, we will perform cross-comparisons between results obtained with two independent sampling codes, parallel Bilby
\cite{Ashton:2018jfp,Smith:2019ucc} and LALInference \cite{Veitch:2014wba}. This is a particularly urgent task, as we expect the number of such atypical events to grow with the improvements in detector sensitivity.
Second, we aim to provide improved parameter estimation results for GW190521 that might be useful to clarify its astrophysical properties. A key result is that we confirm the multi-modal nature of the posterior found in \cite{Nitz:2020mga}, but with some drastic quantitative changes due to improvements in the waveform models, in particular support for both masses being outside the PISN mass-gap, and the support for an intermediate mass ratio binary are drastically reduced with respect to what whas found in \cite{Nitz:2020mga}.

The paper is organized as follows. We first discuss the waveform models we employ in Sec.~\ref{sec:models}, focusing on differences that are relevant for analyzing GW190521, and on how we can test robustness by comparing results from different models.
We then summarize previous results from the literature in Sec.~\ref{sec:previously}.
In Sec.~\ref{sec:PE_setup} we describe our methods for parameter estimation and for checking convergence and consistency between different prior assumptions. Readers interested primarily in new results may wish to skip to our results in Sec.~\ref{sec:results}, which is introduced by a sub-section that briefly outlines 
the types of analyses we have performed and the results we have obtained.
We give our final conclusions in Sec.~\ref{sec:conclusions} and discuss the dependency of the results on spherical harmonic mode content in Appendix \ref{sec:appendixhm}.

\section{Waveforms}\label{sec:models}

\subsection{Notation and conventions}
\label{sec:conventions}

We use the same notation and conventions as in our re-analysis of GWTC-1 \cite{mateulucena2021adding}. 
We will report all masses in units of the solar mass $M_\odot$,
and except where otherwise noted, we always refer to masses in the source frame, assuming a standard cosmology \cite{Planck2015} (see Appendix B of \cite{LIGOScientific:2018mvr}). 
Some figures and tables use a ``$\mathrm{src}$" index as a more explicit notation for clarity.
Individual component masses are denoted by $m_i$,
and the total mass is $M = m_1 + m_2$.
The chirp mass is $\mathcal M = (m_1 \, m_2)^{3/5} M^{-1/5}$.
We define mass ratios $q = m_2 / m_1 \leq 1$ and $Q = m_1 / m_2 \geq 1$.

We also define effective spin parameters which are commonly used in waveform modelling and parameter estimation.
The parameter $\chi_{\rm{eff}}$ is defined as
 \begin{equation}\label{def:chieff}
    \chi_{\mathrm{eff}}=\frac{m_1 \chi_1 + m_2 \chi_2}{m_1 + m_2}, 
 \end{equation}
where the $\chi_i$ are the projection of the spin vectors of the individual black holes onto the instantaneous direction perpendicular to the orbital plane.
The effective spin precession parameter $\chi_\mathrm{p}$ \cite{Schmidt:2014iyl} is designed to capture the dominant effect of precession, and corresponds to an approximate average over many precession cycles of the spin in the precessing orbital plane, and is defined in terms of the average spin magnitude $S_\mathrm{p}$, \cite{Schmidt:2014iyl}
\begin{align}\label{eq:avNNLOSperp}
    S_\mathrm{p} &= \frac{1}{2} \left( A_1 S_{1,\perp} + A_2 S_{2,\perp} + | A_{1} S_{1,\perp} - A_2 S_{2,\perp} |\right) , \\
    &= {\rm{max}} \left( A_1 S_{1,\perp} , A_2 S_{2,\perp} \right),
\end{align}
where $A_1 = 2 + 3 / (2 q)$, and $\chi_\mathrm{p}$
is then defined as
\begin{align}\label{def:chip}
    \chi_\mathrm{p} &= \frac{S_\mathrm{p}}{A_1 m^2_1} .
\end{align}
Both  $\chi_{\mathrm{eff}}$ and $\chi_\mathrm{p}$ are dimensionless and thus independent of the frame (source or detector).

We will employ waveforms with several multipoles beyond the quadrupolar contribution, always considering pairs of both positive and negative modes when referring to a particular multipole. Thereby, to refer to the example list of multipoles $(l,m)=(2,\pm2), (2,\pm1)$, we will use the notation $(l,|m|)=(2,2), (2,1)$ or simply $(2,2),(2,1)$.

\subsection{Waveform models used}
\label{sec:model_descriptions}

\begin{table*}[htpb]
 \caption{Waveform models used in this paper. We indicate which multipoles are
included for each model. For precessing models, the multipoles correspond to those
in the co-precessing frame.
For \phTPHM, we also show comparison results with reduced sets of multipoles at several points in this paper, and in fact we use the $\ell\leq4$ setting as a default run and comparison basis in most studies of alternative model options or priors.}
 \label{tab:models}
 \begin{ruledtabular}
 \begin{tabular}{llcccc}
  \textbf{Family} & \textbf{Full name} & \textbf{Precession} & \textbf{Multipoles $(\ell,\,|m|)$} & \textbf{ref.}\\
\hline
\multirow{1}{*}{SEOBNR} 
   & \seobnrvforphm &  \checkmark & (2, 2), (2, 1), (3, 3), (4, 4), (5, 5) & \cite{Ossokine:2020kjp,Babak:2016tgq,Pan:2013rra}
\\[6pt]
 NR surrogate  & \NRSur &  \checkmark & $\ell \leq 4$ & \cite{Varma:2019csw}\\[6pt]
\multirow{1}{*} {Phenom - Gen. 3}
  & \phPvthreehm  &  \checkmark & (2, 2), (2, 1), (3, 3), (3, 2), (4, 4), (4, 3) & \cite{Khan:2019kot}
\\[6pt]
\multirow{2}{*}{PhenomX}  &
\phXHM &  $\times$ & (2, 2), (2, 1), (3, 3), (3, 2), (4, 4)  & \cite{Garcia-Quiros:2020qpx,Garcia-Quiros:2020qlt} \\
  & \phXPHM &  \checkmark & (2, 2), (2, 1), (3, 3), (3, 2), (4, 4) & \cite{Pratten:2020ceb} &\\[6pt]
\multirow{2}{*}{PhenomT} 
   & \phTHM &  $\times$ & (2, 2), (2, 1), (3, 3), (4, 4), (5, 5)  &  \cite{Estelles:2020osj,Estelles:2020twz}\\
  & \phTPHM &  \checkmark & (2, 2), (2, 1), (3, 3), (4, 4), (5, 5) & \cite{Estelles:2020osj,estelles2021new}
 \end{tabular}
 \end{ruledtabular}
\end{table*}

For a complete list of all the waveform models used in the present paper see Table \ref{tab:models}. 
The original LVC publications on GW190512 \cite{Abbott:2020tfl,Abbott:2020mjq} used three families of waveform models, which represent
incarnations of three well established approaches to CBC waveform modelling:
\begin{itemize}
\item The \seobnrvforphm time-domain model \cite{Ossokine:2020kjp,Babak:2016tgq, Pan:2013rra}: constructed within the effective-one-body (EOB) framework \cite{Akcay:2018yyh,Bohe:2016gbl,Cotesta:2020qhw,Damour:2016bks,Damour:2015isa,Damour:2012ky,Nagar:2018plt,Nagar:2018zoe,Nagar:2018gnk,Nagar:2017jdw,Nagar:2016iwa,Nagar:2015xqa,Purrer:2015tud,Taracchini:2012ig,Taracchini:2013rva,Damour:2001tu,Cotesta:2018fcv}.
\item The \NRSur surrogate time-domain model \cite{Varma:2019csw}, which directly interpolates NR waveforms.
\item The \phPvthreehm frequency-domain model, which corresponds to the third generation
of models in the \ph family \cite{Husa:2015iqa,Khan:2015jqa,Hannam:2013oca,Bohe:PPv2,London:2017bcn,Khan:2018fmp,Khan:2019kot}.
\end{itemize}
Here we employ two further recently developed models that represent upgrades of \phPvthreehm and constitute a fourth generation of \phIMR models:
\begin{itemize}
\item The frequency-domain model \phXPHM \cite{Pratten:2020ceb} which builds upon an underlying non-precessing model \phXHM \cite{Pratten:2020fqn,Garcia-Quiros:2020qpx,Garcia-Quiros:2020qlt} that features calibration of the subdominant harmonics to NR simulations.
\item \phTPHM \cite{Estelles:2020osj,estelles2021new}, building on the non-precessing model \phTHM \cite{Estelles:2020osj, Estelles:2020twz}, which essentially applies the same phenomenological techniques at the heart of \phXPHM to construct a native time-domain model. Working in the time domain allows several key improvements that we will discuss below.
\end{itemize}

All these models include a description of precession effects and sub-dominant harmonics, but do not include eccentricity, and they have complementary strengths and shortcomings that we will detail below.

Only \NRSur is calibrated to precessing NR waveforms, but its training dataset is restricted to mass ratio $Q\leq 4$ and dimensionless component spin magnitudes $a_{\mathrm{1,2}}\leq 0.8$). However the model can also be evaluated in the  extrapolation regions region with $Q \leq 6$ and $a_{\mathrm{1,2}}\leq 1$. Furthermore, usage of the model is restricted by the length of the original time-domain NR waveforms, and restrictions get tighter in the frequency domain due to the need of windowing before Fourier transforming the template. The limited length of \NRSur waveforms leads, in particular, to extra constraints on the minimum frequency and total mass allowed in parameter estimation analyses.

The \ph models describe precession via an approximate map between signals from non-precessing and precessing systems, which we will refer to as ``twisting-up'' \cite{Schmidt:2010it,Schmidt:2012rh,Hannam:2013pra}. This approximation exploits the fact that, at least during the inspiral, the precession time scale is much slower than the orbital time scale, and thus the precessing motion mainly acts as an amplitude modulation. The spin of the remnant of the precessing system is however in general significantly different from the final spin of the non-precessing system. For a recent discussion of the approximations of this approach see \cite{Ramos-Buades:2020noq}.

The  \seobnrvforphm  model \cite{Buonanno:2002fy,Babak:2016tgq,Pan:2013rra,Cotesta:2018fcv,Ossokine:2020kjp} numerically integrates in the time domain the EOB BBH dynamics, including the spin-precession equations, using a Hamiltonian and GW flux that are tuned to non-precessing NR simulations. Then, the waveforms in the inertial frame are obtained by applying a time-domain rotation (“twisting-up”) to the waveforms in the co-precessing frame \cite{Buonanno:2002fy,Babak:2016tgq,Pan:2013rra}. 

Neither the \seobnrvforphm nor \phIMR models include the asymmetries between positive and negative $m$
spherical harmonic modes in the co-precessing frame, which are related to the large recoil velocities observed in some NR simulations of precessing binaries, as discussed in \cite{Brugmann:2007zj}.

We now turn to describe relevant aspects of the \phXF and \phT families which we use for the new results presented in this paper.
Frequency-domain waveform models are particularly attractive for GW data analysis, since they are naturally adapted to a matched-filter type analysis, where the noise is characterized in the frequency domain, and accordingly allow the most computationally efficient Bayesian inference analysis. In order to accelerate the evaluation of precessing waveforms, current frequency domain \phIMR  models also use the SPA approximation to compute the transfer functions between the frequency-domain non-precessing waveform and the precessing waveform in an inertial frame (see \cite{Marsat:2018oam} for more accurate alternatives).
The assumptions underlying the SPA fail for merger and ringdown, but the method has been found to work surprisingly well, and has been routinely used in GW data analysis, see e.g. \cite{LIGOScientific:2018mvr,Abbott:2020niy}. The approximation does however have to be employed with caution when essentially the only observable part of the signal is the merger and ringdown, as is the case for GW190521. This shortcoming has been one of the main reasons for us to also develop a time-domain phenomenological waveform model, \phTPHM, which does not rely on the SPA. One of the goals of this paper is to discuss in detail how to avoid misleading conclusions from analysing GW190521 with \phXPHM, and how improved results can be obtained with \phTPHM.

Since neither \phXPHM nor \phTPHM are calibrated to precessing NR waveforms but rather build on the above approximations to describe precession effects, it is essential to incorporate in the models some functionality to test the robustness of results for challenging events like GW190521. As  discussed in Appendix F of \cite{Pratten:2020ceb} for \phXPHM  and in the appendix of \cite{estelles2021new} for \phTPHM, the LALSuite \cite{lalsuite} implementation of our models supports several options regarding the choice of precession prescription and final spin approximations. 
These options are selected with parameters that take integer values, which we will refer to as $\texttt{PV}$ for the inspiral precession version and $\texttt{FS}$ for final spin.

Our ``twisting-up'' procedure is based on time/frequency dependent rotations from the co-precessing frame to an inertial one in which we observe the signal. For the \ph models this rotation is implemented through three Euler angles. \phPvtwo only supports an effective single-spin, orbital-averaged description valid at next-to-next-to-leading  (NNLO) post-Newtonian order.
\phXPHM allows using the same prescription, but as a default  
it relies on a more recent double-spin description that can be derived within the post-Newtonian framework using multiple-scale-analysis (MSA) \cite{Chatziioannou:2017tdw} (this description is also used by \phPvthreehm). 
\phTPHM implements both NNLO and MSA Euler angles, but its default behavior is to numerically integrate evolution equations for the component spins as discussed in \cite{estelles2021new}. We will refer to different precession prescriptions with the acronym \texttt{PV} and follow the same convention enforced in LALSuite, where \texttt{PV}=223 corresponds to the MSA approximation and \texttt{PV}=300 to the numerically integrated angles. 

Another setting of the models that can be specified by the user is the final spin approximation, as discussed in Sec.~IV.D of \cite{Pratten:2020ceb} for \phXPHM and Sec.~II.E of \cite{estelles2021new} for \phTPHM. 
The default choice for \phXPHM is to use a precession-averaged equation inspired by the MSA formalism. This version will be referred to as version {\tt FS}=3. Alternative versions attach the in-plane spins to the larger mass, either relying on the usual effective precession spin $\chi_\mathrm{p}$ ({\tt FS}=0, which is adopted by all third-generation \phIMR models), or by taking the norm of the in-plane spin vectors at the reference frequency ({\tt FS}=2). The default version of \phTPHM takes the norm of the in-plane spin vectors at the coalescence time ({\tt FS}=4).

There are several improvements in the treatment of precession achieved by the time-domain \phTPHM in comparison with the frequency-domain \phXPHM. 
First, in order to obtain explicit expressions for the spherical harmonic modes of the precessing frequency-domain models, \phXPHM and previous \phIMR models use the SPA to compute approximate Fourier transforms. 
Second, in the time domain it is simple to incorporate analytical knowledge about the ringdown frequencies in the ringdown portion of a precessing waveform, see Sec.~II.E of \cite{estelles2021new}. This has not yet been achieved in the frequency domain. This is particularly crucial for GW190521, where a large part of the observed signal-to-noise ratio (SNR) is due to the ringdown portion of the signal.
Third, the numerical integration of the equations for the spin dynamics in \phTPHM also resolves an inconsistency of the MSA Euler angles with the non-precessing limit, which we discuss in \cite{estelles2021new}. The numerical integration also provides further gains in accuracy and we find it to decrease the computational cost \cite{estelles2021new}.

Finally, we note that a particularly challenging region of the BBH parameter space arises at larger mass ratios, where the precession cone angle can become large, and the orbital angular momentum ${\mathbf L}$ can become smaller than the (sum of the) component spins. Then both ${\mathbf L}$ and the total angular momentum ${\mathbf J}$ may end up flipping their direction. The latter situation is also known as transitional precession \cite{Apostolatos:1994mx, Kidder:1995zr, Schmidt:2012rh}, as opposed to simple precession, when ${\mathbf J}$ at least approximately maintains its direction. Very few NR simulations exist for the cases of large angles between ${\mathbf J}$ and ${\mathbf L}$, and for the sign of ${\mathbf J}$ flipping, and these situations are related to various caveats in the post-Newtonian and MSA approximations that are often used in waveform modelling, and in particular in the \phIMR models. While the systematic errors of precessing waveform models are in general not yet very well understood, this is particularly true when the normal to the orbital plane or the final spin flip sign with respect to the direction at large separation, This situation indeed arises for  the results of \cite{Nitz:2020mga}.
In such cases one should proceed with great caution, and test the robustness of results by comparing different waveform models. Future work will aim to improve the robustness of our models for such situations.

\section{Summary of previous results}
\label{sec:previously}

\begin{figure*}[htpb!]
\includegraphics[width=\columnwidth]{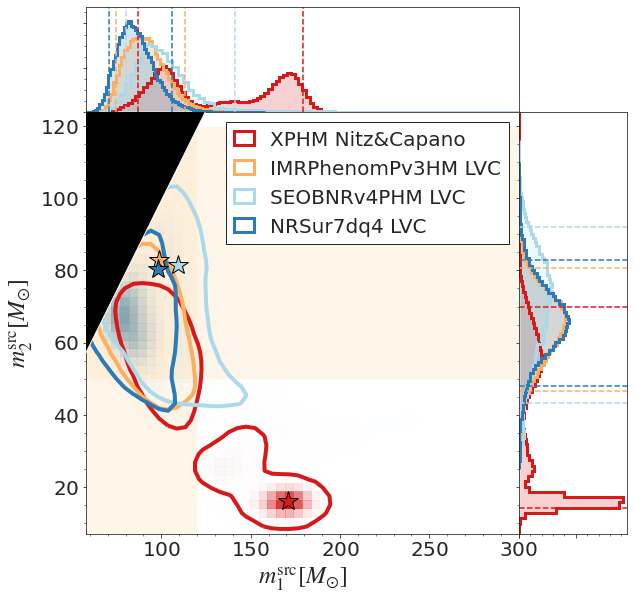}
\includegraphics[width=\columnwidth]{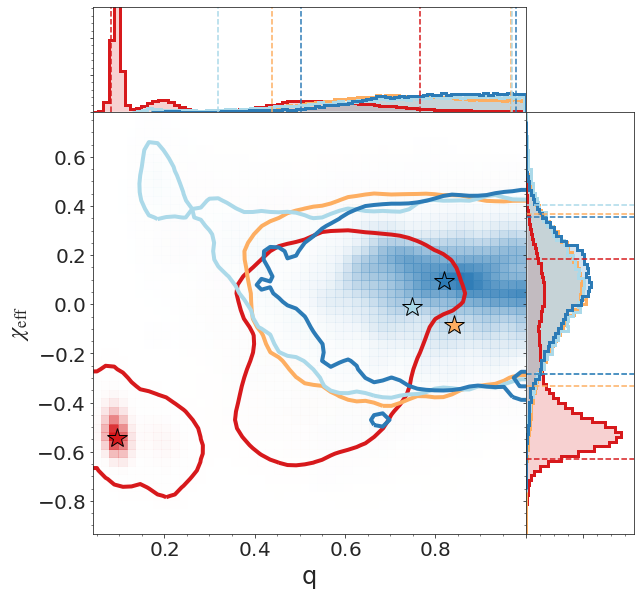}
 \caption{Comparison of inferred posterior distributions for the official results from the LVC~\cite{Abbott:2020tfl,Abbott:2020mjq} and the results from Nitz and Capano~\cite{Nitz:2020mga} (the latter have been re-weighted to a flat in component mass prior, in the detector frame).
 Here and in similar figures throughout the paper, the central panel shows the 2D joint posteriors with contours marking 90\% credible intervals, while the smaller panels on top and to the right show the corresponding 1D distributions for the individual parameters, with the 90\% credible interval indicated by the dashed lines. Plot ranges account for all posterior samples, unless a specific range is specified. 
 The $\maxL$ values from the posterior samples of each run are highlighted as stars in the central panels.}
\label{fig:lvcNCcomparison}
\end{figure*}

Due to its exceptional nature, GW190521 was the subject of two dedicated LVC publications \cite{Abbott:2020tfl,Abbott:2020mjq}; later it was also reanalyzed in the context of GWTC-2 \cite{Abbott:2020niy}. Only results obtained with \NRSur are shown in the discovery paper~\cite{Abbott:2020tfl}, while \cite{Abbott:2020mjq} also presents results obtained with \seobnrvforphm and \phPvthreehm.
The mass-ratio prior in these LVC analyses was constrained to $q\geq 0.17$, matching the \NRSur extrapolation region.
They also used a flat prior in detector-frame masses and a power-law $d_{L}^2$ distance prior (albeit the latter was changed to a uniform in comoving source-frame volume prior in \cite{Abbott:2020niy}). GW190521 was among the few events in GWTC-2 for which spin magnitudes could be constrained to be non-zero: this is also reflected in its relatively large inferred $\chi_\mathrm{p}$ ($\approx$ 0.7 median value). The estimated mass ratio when running with \NRSur was $q=0.79_{-0.29}^{+0.19}$ and runs performed with the other two waveform approximants delivered very similar results. The LVC analysis has strong astrophysical implications, as it places either or both components in the pair-instability supernova mass-gap (PISN) and the final remnant in the realm of intermediate-mass black holes, for which no conclusive evidence existed at the time of publication. 
Note that the limits of the PISN mass gap are uncertain, they have been placed at approximately 50 and 130 solar masses in \cite{Woosley:2016hmi}, but a more recent analysis \cite{Woosley:2021xba} suggests the lower limit could be as high as 70 solar masses, and the upper limit as high as 161 solar masses. Another recent analysis \cite{mehta2021observing} placed the limits at 59 and 139 solar masses. For the LVC analysis lower limits of 50 and 65 solar masses were employed.
Fishbach and Holz \cite{Fishbach:2020qag} challenged this conclusion starting from the observation that the merger-rate of systems involving a mass-gap component is expected to be very low. By imposing a population-informed prior, they concluded that GW190521 can be considered a ``straddling" binary, where neither component can be confidently placed within the mass-gap. In particular, they find that, under the assumption that the secondary mass falls below the mass gap, then the primary mass distribution has a large support above the mass gap. This conclusion was supported in a later study conducted by Nitz and Capano \cite{Nitz:2020mga}, who suggested that the relatively tight constraint on the mass ratio imposed by the LVC analysis, coupled with the choice of luminosity distance prior and sampler settings (among which insufficient live points), led initial parameter estimation studies to exclude the highest likelihood region for this event. 
A comparison of results from the LVC and Nitz--Capano results is shown in Fig.~\ref{fig:lvcNCcomparison}, based on the publicly available posterior data.
Their reanalysis
also identifies the primary as an IMBH, with a mass confidently above $100\,M_{\odot}$.
The authors explored the impact of different mass priors (in the source frame) and imposed a uniform in comoving-volume prior on the luminosity distance, running both \NRSur and the more recent \phXPHM, which had only passed internal LVC review and become publicly available as part of LALSuite \cite{lalsuite} on April 1 2020 and hence was not yet included in \cite{Abbott:2020tfl,Abbott:2020mjq}. They found strong support for very unequal masses ($q\leq 0.25$) and clear signs of multi-modalities in the posteriors, which could not be eliminated when re-weighting to match the LVC priors. In particular, three distinct modes were identified, with $Q\approx1, 5, 10$. The support for the $Q\approx 10$ mode was enhanced when using a flat prior in $Q$ which favors more unequal-mass systems.
Their analysis however did not yet use the latest version of the \phXPHM model nor explored different options of this model, and the default version at that point used a prescription for the final spin of the merger remnant that has since been updated in the publicly available LALSuite version~\cite{lalsuite}.

In a more recent paper~\cite{Capano:2021etf}, Capano et al. have analyzed GW190521 with model-agnostic ringdown signals, extracting an additional mass ratio measurement from only the ringdown part of the signal.
The resulting posterior is unimodal, peaked away from equal masses, but broadly consistent with both the LVC results and the two lower-$Q$ peaks of their previous results.
They also include a re-run of their \phXPHM analysis with the updated default final-spin prescription which we discuss in Secs.~\ref{sec:model_descriptions} and \ref{sec:results_xphm} of this present paper and in more detail in the recently updated Sec. IV D of \cite{Pratten:2020ceb}.
Those updated results no longer support the third mode at $Q\approx10$ and are overall consistent with the results we present here, and with another run with the updated \phXPHM default version presented in parallel by \cite{mehta2021observing}. These results have been released together with a  re-analysis of the public LIGO-Virgo data from the O1, O2 and O3a observing runs \cite{nitz20213ogc}, and are available in a companion data release for \cite{nitz20213ogc}. We have compared the results presented here for \phXPHM with the updated results from \cite{nitz20213ogc}, finding broad consistency, but with some differences in the recovered posteriors. A comparison of low-mass events from our re-analysis of GWTC-1 \cite{mateulucena2021adding} with their results reported for these events shows larger discrepancies, as we discuss in \cite{mateulucena2021adding}. The good agreement we find here and in \cite{mateulucena2021adding} between different parameter estimation codes, LALInference \cite{Veitch:2014wba} and parallel Bilby \cite{Smith:2019ucc}, suggests that our results are robust. Tracking down the reasons for the differences found with respect to the results reported in \cite{nitz20213ogc} would require further work; we do note however that \cite{nitz20213ogc} use a different estimate of the noise power spectral density (PSD), and to our knowledge no calibration uncertainty estimates are employed.

\section{Methodology for parameter estimation}\label{sec:PE_setup}

\subsection{Data set}\label{sec:data}

We use public GW strain data 
collected by the Advanced LIGO detectors \cite{TheLIGOScientific:2014jea} and Advanced Virgo detector \cite{TheVirgo:2014hva} from the Gravitational Wave Open Science Center (GWOSC) \cite{Vallisneri:2014vxa,Abbott:2019ebz}, as well as PSDs and calibration uncertainties included in the GWOSC release  \cite{GW190521:datadcc}. 
From the available GWOSC strain data sets we have selected the data sampled at 16\,kHz, with a sampling rate of 1024\,Hz chosen for our analysis, consistent with the choice in \cite{Abbott:2020tfl,Abbott:2020mjq}.
The lower and upper cutoff frequencies for the likelihood integration were taken to be 11.0\,Hz and 512\,Hz (the Nyquist frequency corresponding to the sampling rate), again consistent with \cite{Abbott:2020tfl,Abbott:2020mjq}.

\subsection{Sampling codes}\label{sec:samplers}

We have carried out Bayesian parameter estimation of the signal using two publicly available codes, the \texttt{Python}-based parallel Bilby (\texttt{pBilby}, PB) code  \cite{Ashton:2018jfp,Smith:2019ucc}, which uses the {\tt dynesty} \cite{10.1093/mnras/staa278} variant of the nested sampling algorithm \cite{Skilling:2004ns}, and the \texttt{LALInference} (\texttt{LI})
code \cite{Veitch:2014wba},
which is part of the LALSuite \cite{lalsuite} package for GW data analysis, using its implementation of Markov Chain Monte Carlo (MCMC) sampling.

Parallel Bilby provides a highly parallel and flexible implementation of nested sampling, and supports a range of priors and choices of sampling parameters and settings.
With \texttt{pBilby}, we sample in mass ratio and chirp mass, which is easier than sampling the component masses in that code.
We largely use the default settings of the code apart from the following choices: we fix the minimal ({\tt walks}) and maximal ({\tt maxmcmc}) number of MCMC steps to 200 and 15000 respectively.
For our final results we have set the number of autocorrelation times to use before accepting a point ({\tt nact}) to a value of 30. We have varied the number of nested sampling live points ({\tt nlive}) between 1024 and 4096 for selected runs to test that we have obtained (sufficiently) converged results,
and always show the results for {\tt nlive}$=4096$.
In order to speed up calculations we use distance marginalization as described in \cite{Thrane_2019}. For each of the \texttt{pBilby} runs we quote results for, we have carried out four independent simulations (independent seeds), and then merged the posteriors to a single posterior with PESummary \cite{Hoy:2020vys}. 

LALInference samples in mass ratio and chirp mass, re-weighting to a prior that is flat in component masses as described in \cite{Veitch:2014wba}. We use essentially standard LALInference settings with eight temperatures, but a large number of independent chains, 120 for our production runs. For our LALInference runs we do not employ the distance marginalization used for our Bilby runs.

We have previously used \texttt{pBilby} as our primary code for our re-analysis of the GW190412 event \cite{Colleoni:2020tgc,estelles2021new,mateulucena2021adding}, where we found good agreement with LALInference results as reported in \cite{Colleoni:2020tgc}. We have however found that the computational cost of comparably well sampled \texttt{pBilby} runs is significantly higher than for LALInference runs due to the high required settings of the {\tt nact} parameter.
Here we use LALInference for our primary results and pBilby for comparisons.

\subsection{Priors}\label{sec:priors}

Runs performed with \texttt{pBilby} have been sampled with a prior uniform in ``inverse'' mass ratio $Q=1/q$, following \cite{Nitz:2020mga}. This prior emphasizes unequal masses and improves \texttt{pBilby} convergence in the unequal-mass regime. Unlike \cite{Nitz:2020mga}, we choose to sample in the detector-frame, to take advantage of distance-marginalization, which would require a non-standard likelihood in Bilby if sampling in source-frame. In most of the results shown, and as indicated  when stating results,
we have performed a post-processing re-weighting from this prior to a prior flat in component masses with the corresponding functions from the Bilby code, to obtain results matching the same prior as the LALInference runs (flat in component masses).

Additionally we have performed some runs with restricted versions of these priors to improve resolution for more unequal masses. In particular, we report in Sec. \ref{sec:results} results for LALInference runs in a mass ratio range $q\in[0.035,0.15]$ and \texttt{pBilby} runs in a range $q\in[0.035,0.2]$. For studying the possible association of the event with the reported AGN flare \cite{Graham:2020gwr} we have also performed runs fixing the sky location and luminosity distance to the values reported for the AGN flare.

Finally, we have employed mainly three different sets of mass prior ranges for the runs reported in this work, checking that we avoid any significant railing against prior limits. Note that a small amount of railing against the lower mass-ratio bound is still present; however we decided to not attempt to precisely map out the tail at low $q$ as models become less reliable and computational cost increases significantly. We comment on the possibility of further low-$q$ posterior modes in Sec.~\ref{sec:conclusions}.
For \phTPHM and \phTHM runs with LALInference, we have employed a prior uniform in component masses (in detector frame) with a range of $m^\mathrm{det}_{1,2}\in[10,400]M_{\odot}$ and default mass ratio constraints $q\in[0.035,1.0]$. For \phTPHM runs with \texttt{pBilby} we have employed a prior uniform in inverse mass ratio with range $q\in[0.035,1.0]$ and uniform in total mass with range $M^\mathrm{det}_\mathrm{T}\in[80,550]M_{\odot}$, and constraints for component masses (in detector frame) $m^\mathrm{det}_1\in[30,400]M_{\odot}$ and $m^\mathrm{det}_2\in[5,400]M_{\odot}$.
For \phXHM and \phXPHM runs, we have employed the same ranges in LALInference as with \phTPHM runs, but for \texttt{pBilby} runs the parameter ranges are: $q\in[0.04,1.0]$, $m^\mathrm{det}_1\in[30,300]M_{\odot}$, $m^\mathrm{det}_2\in[5,200]M_{\odot}$ and $M^\mathrm{det}_\mathrm{T}\in[80,550]M_{\odot}$. Differences in the mass ranges are due to problems found with railing posteriors for the \phTPHM runs, which in general have support for higher component masses.

In some cases we also reweight the posterior samples obtained with a certain set of priors to obtain an approximation of what the posterior should be when using a different set of priors, without having to run the full alternative inference. Specifically, we use the bilby implementation of this reweighting procedure which converts samples with a given prior in chirp mass and mass-ratio to a prior flat in component masses by resampling the posterior with weights defined as the ratio in new-over-old prior values times the Jacobian of the transformation. The procedure produces new posteriors that contain only $25\%$ of the original number of samples.

All runs have maximum component spin magnitudes limited to 0.99, and the luminosity distance prior is chosen as uniform in comoving volume, assuming the Planck15 \cite{Planck2015} cosmology with a range $D_L\in[0.2,10]\,\text{Gpc}$. The LALInference prior contains an additional factor $1/(1+z_{L})$, accounting for time dilation, which is not present in the definition for \texttt{pBilby}; but using the reweighting procedure on \texttt{pBilby} results, we have tested that this does not have any noticeable influence on estimates of $D_L$ or any other quantities.

\subsection{Maximum likelihood values and waveforms}\label{sec:maxL}

The main results of Bayesian parameter estimation are the posterior distributions, and point estimates are usually given as medians with error estimates given by the 90\% confidence intervals.
Sometimes, it can also be enlightening to consider the maximum-likelihood value ($\maxL$) returned by an analysis, and which point in parameter space it correspond to, as the likelihood directly answers the question of how well the employed model (across the sampled part of parameter space) can fit the data.
However, there are several caveats in interpreting the $\maxL$ values over a set of posterior samples, since a Bayesian parameter estimation run, such as those we employ here using \texttt{pBilby} and LALInference, is by construction not an optimal $\maxL$ finding algorithm. The prior has significant influence on how densely which parts of the parameter space are evaluated, and the $\maxL$ reported over the final posterior sample may be far from the actual maximum over all likelihoods evaluated while the sampling chains progressed.
It is important to note that the number of posterior samples is typically much smaller than the number of likelihood evaluations, and to achieve a good estimate of $\maxL$ much more expensive sampling settings may be required than in order to get good estimates for source parameter values and error estimates.
Nevertheless, comparing the $\maxL$ across runs can be a useful additional diagnostic of the behavior of waveforms and samplers, and we highlight them on all posterior plots in this paper.

Indeed for GW190521, we find that the maximum-likelihood parameters do not appear to be stable across runs, and are influenced by statistical fluctuations, sampler settings, as well as waveform models and priors.
For an example, see Sec.~\ref{sec:results_xphm}.

\section{Results}
\label{sec:results}

\subsection{Overview}

\begin{figure*}[htpb!]
\includegraphics[width=\columnwidth]{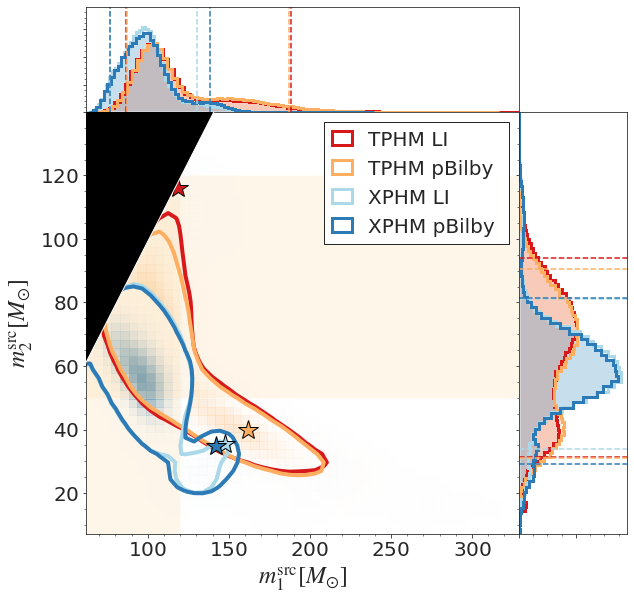}
\includegraphics[width=\columnwidth]{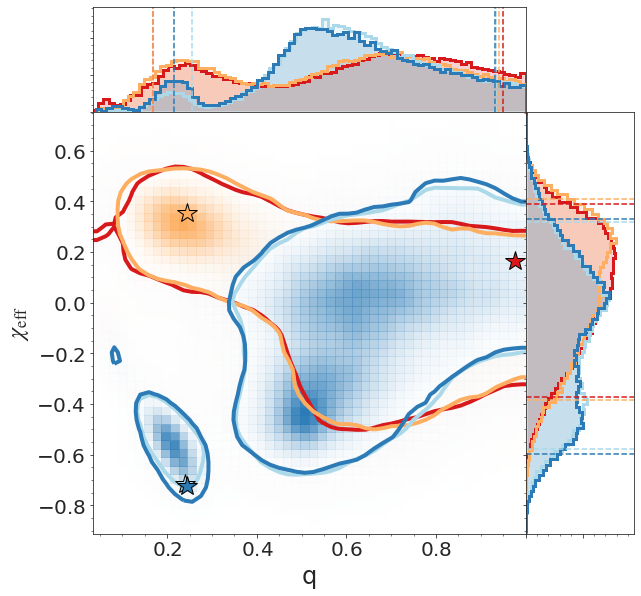}
\caption{Two-dimensional joint posterior distributions for source-frame masses (left panel), and mass ratio and effective spin (right panel) obtained with the default versions of \phTPHM (red: LALInference, orange: pBilby) and \phXPHM (light blue: LALInference, dark blue: pBilby). Dashed vertical lines in the one-dimensional plots mark 90\% confidence intervals and stars mark the $\maxL$ values.
Unless otherwise indicated, here and in the following figures and tables \phTPHM results correspond to $\ellmax=4$.
\label{fig:PBLIcomparison_mq}
}
\end{figure*}

\begin{figure*}[tpbh!]
\includegraphics[width=\columnwidth]{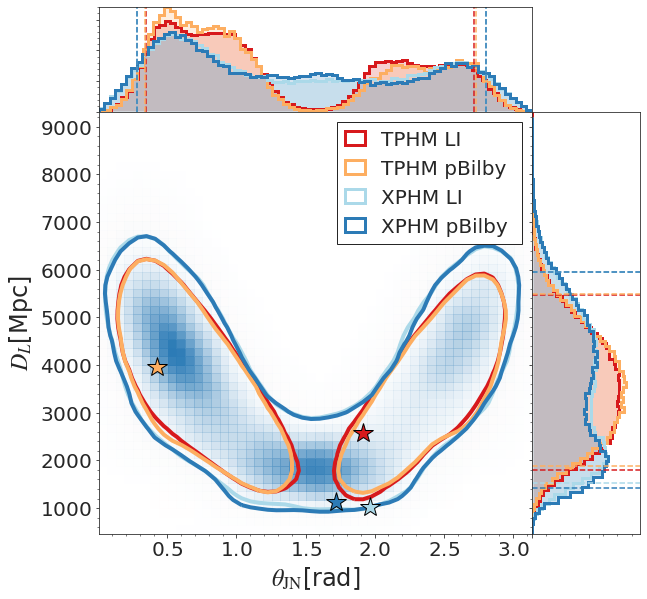}
\includegraphics[width=\columnwidth]{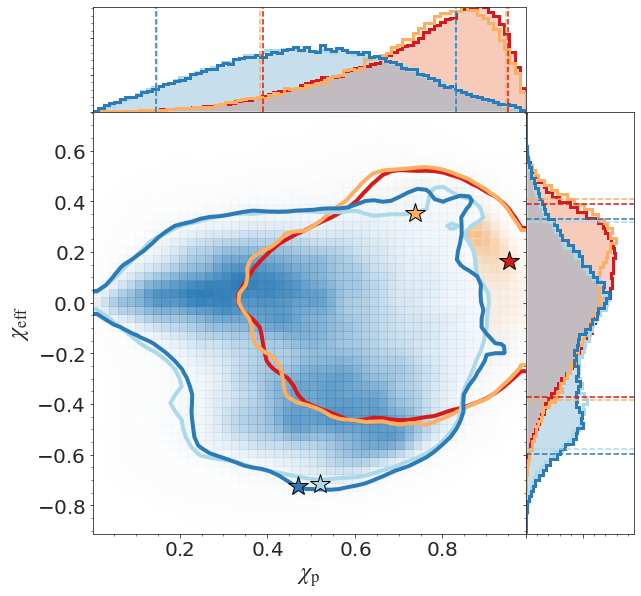}
\caption{Two-dimensional joint posterior distributions for distance and inclination (left panel), as well as for effective and precession spin parameters (right panel), obtained with the default versions of \phTPHM (red: LALInference, orange: pBilby) and \phXPHM (light blue: LALInference, dark blue: pBilby). Dashed vertical lines in the one-dimensional plots mark 90\% confidence intervals and stars mark the $\maxL$ values.
\label{fig:PBLIcomparison_spinLoc}
}
\end{figure*}

\begin{figure}[htpb!]
\includegraphics[width=\columnwidth]{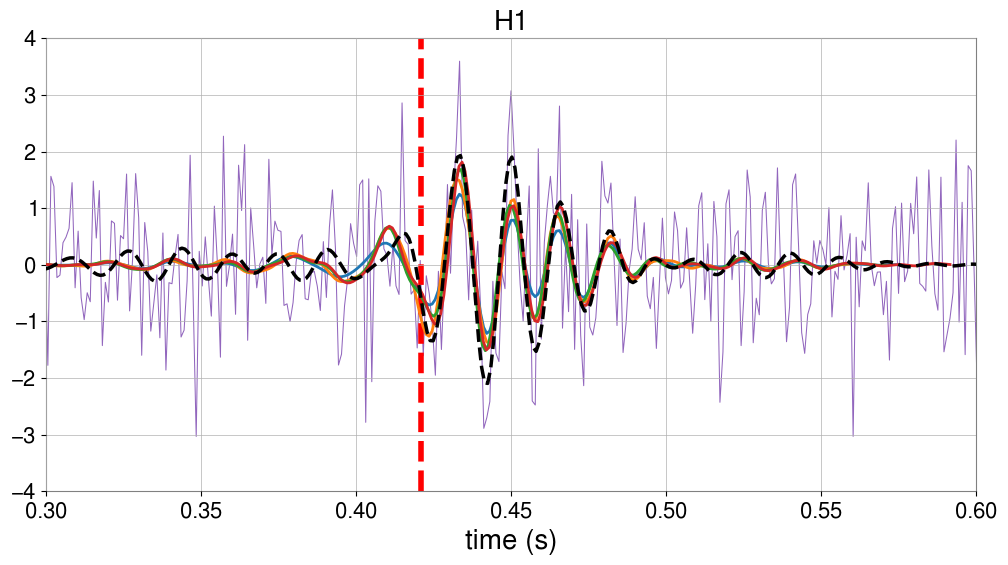}
\includegraphics[width=\columnwidth]{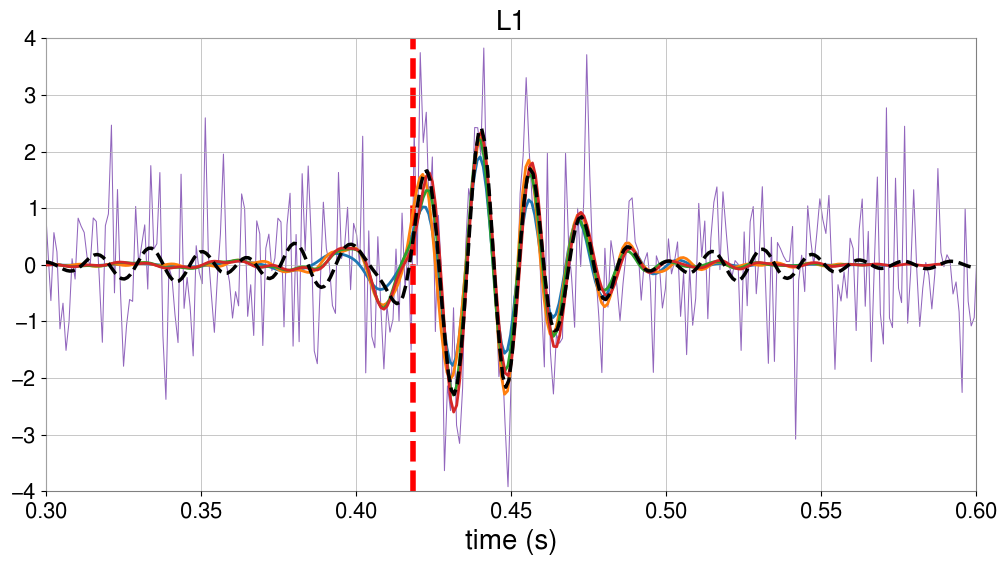}
\includegraphics[width=\columnwidth]{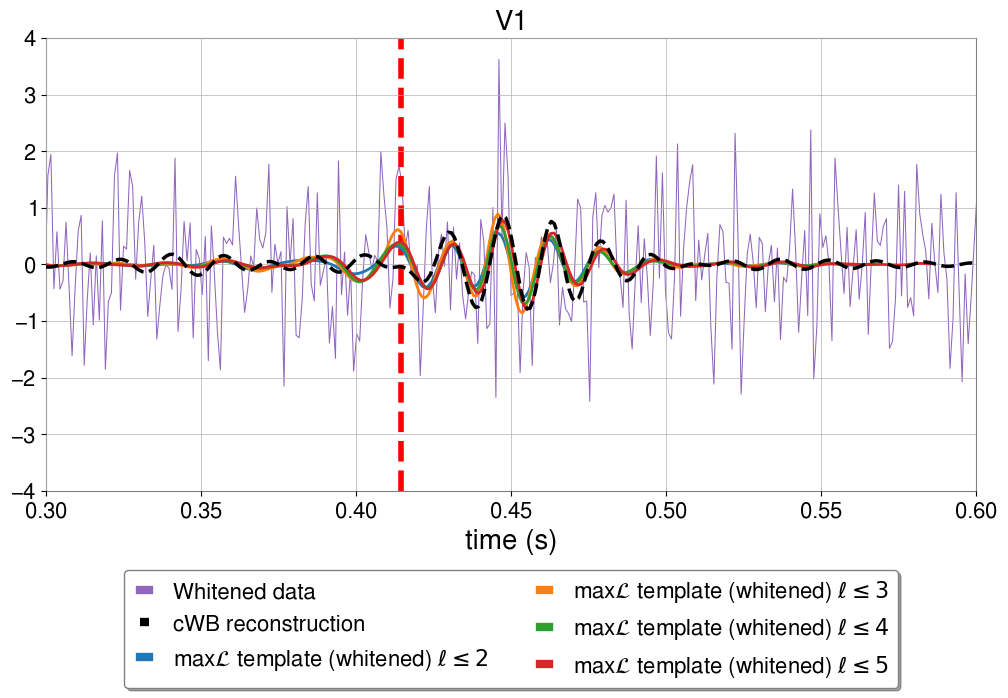}
\caption{Comparison of the maximum-likelihood ($\maxL$) templates from LALInference runs with the \phTPHM PV=300 FS=4 model against detector data for different harmonic content indicated by $\ellmax$ from 2 to 5.
Each panel shows the time-domain detector data of LIGO Hanford (H1), LIGO Livingston (L1) and Virgo (V1) respectively, after whitening by the instrument's noise amplitude spectral density (purple lines), along with point estimate waveform reconstructions from the cWB analysis (dashed black lines, from~\cite{GW190521:datadcc})
and the \phTPHM $\maxL$ templates whitened by the instrument's noise amplitude spectral density (colored solid lines). Red dashed vertical lines show the coalescence time as estimated with \phTPHM. Times shown are relative to May 21, 2019 at 03:02:29 UTC.}
\label{fig:waveforms}
\end{figure}

Our main results derive from the posterior distributions we have obtained with the \phXPHM and \phTPHM models. To test the influence of the harmonic content in the templates, we will in general present results for different harmonic content for \phTPHM. In Figs. 
\ref{fig:PBLIcomparison_mq} and \ref{fig:PBLIcomparison_spinLoc} we show these posteriors for some key parameters: the component masses, mass ratio, effective spins $\chieff$ and $\chip$, luminosity distance $d_L$, and the angle $\theta_{JN}$ between ${\mathbf J}$ and the line of sight. We find consistency between the results obtained with LALInference and pBilby after re-weighting the pBilby results (with the uniform-in-$Q$ prior) to a prior that is flat in component masses, giving us confidence in our sampling of parameter space.
Recovered SNRs and signal-versus-noise Bayes factors for our main runs and several different model options are shown in Table~\ref{tab:SNR_BF_table}.
Point estimates for key parameters of the main runs are also summarized in Table~\ref{tab:parameters}, again comparing with those from \cite{Nitz:2020mga,Abbott:2020tfl,Abbott:2020mjq}. Complete posterior datasets for our standard LALInference \phTPHM run with $\ellmax=4$ and standard \phXPHM run can be found in our Zenodo data release \cite{hector_estelles_2021_4751655}.

To demonstrate the overall behavior of the GW190521 signal and the quality of the match with our waveform models, in Fig.~\ref{fig:waveforms} we show the $\maxL$ templates from several \phTPHM runs with different mode content compared to the whitened detector data at the time of GW190521 and a waveform reconstruction from the unmodelled cWB analysis~\cite{Klimenko:2015ypf,Abbott:2020tfl,Abbott:2020mjq}.
The detector data is best matched by the ringdown region of the \phTPHM model, while the cycles before merger are suppressed once whitened by the instrument's noise amplitude spectral density.

Below we will discuss details and consequences of the posterior results shown in Figs.~\ref{fig:PBLIcomparison_mq} and \ref{fig:PBLIcomparison_spinLoc}, and we will analyze further posterior distributions, starting with different versions of the non-precessing versions of our models in Sec.~\ref{sec:nonprec}, where we find good agreement between them. This serves as the more solid basis for the more challenging analysis with precessing models.
We then use the \phXPHM frequency-domain model in Sec.~\ref{sec:results_xphm} and compare with the analysis of Nitz and Capano \cite{Nitz:2020mga}, and discuss effects of a code change we have implemented in the default version of \phXPHM, tracking the flipping of direction of the total angular momentum $J$ in the same way as for non-default versions.
We then investigate the case for multimodality in the mass parameters reported by~\cite{Nitz:2020mga} in Sec.~\ref{sec:results_multimode}, showing that we recover a multimodal mass posterior both with the \phXPHM and \phTPHM models,
although with modified details compared to
\cite{Nitz:2020mga}.
Then in Sec.~\ref{sec:results_prec} we investigate the support for precession in the source system, comparing results obtained with precessing and non-precessing approximants. Finally we study the implications for component masses in the mass gap and the support for the association with the AGN flare as a possible electromagnetic counterpart in Secs.~\ref{sec:results_gap} and \ref{sec:results_EM}. Further analysis of the importance of the multimode harmonic content is presented in appendix \ref{sec:appendixhm}.
A crucial part of this analysis is to build confidence in our results by showing consistency between results obtained with different priors and different samplers (nested sampling \cite{Skilling:2004ns} as implemented in pBilby and MCMC as available through LALInference).

\begin{table*}[htpb]
\caption{Network matched-filter SNRs with 90\% credible intervals and $\log$ signal-to-noise Bayes factors $\mathcal{BF}$ for runs with waveform models in the \phenX and \phT families, including several different options of the \phTPHM model.
We note that the highest $\mathcal{BF}$ values are recovered by \phTPHM with reduced mode content. This is consistent with the slightly negative Bayes factor for dominant vs. higher modes reported for the \NRSur model in \cite{Abbott:2020tfl,Abbott:2020mjq}, but as discussed in appendix~\ref{sec:appendixhm} the posteriors become much better resolved once including modes up to $\ell\leq4$, and seem mostly converged in comparison to adding further modes $\ell\leq5$, and hence we use the $\ell\leq4$ run as our main result in this paper.}
\label{tab:SNR_BF_table}
\begin{ruledtabular}
\begin{tabular}{lccccc} 
 Approx. & $\log \mathcal{BF}$ & $\rho_{\mathrm{mf}}^{\mathrm{H}}$ & $\rho_{\mathrm{mf}}^{\mathrm{L}}$ & $\rho_{\mathrm{mf}}^{\mathrm{V}}$ & $\rho_{\mathrm{mf}}^{\mathrm{N}}$  \\ \hline
 XHM  & $80.06\pm0.15$ & $8.0^{+0.2}_{-0.3}$ & $11.8^{+0.5}_{-0.3}$ & $2.4^{+0.7}_{-1.2}$ & $14.4^{+0.3}_{-0.3}$\\
 \rule{0pt}{3ex}%
 XPHM PV=223 FS=3 & $80.43\pm0.21$ & $7.9^{+0.2}_{-0.3}$ & $11.8^{+0.5}_{-0.3}$ & $2.5^{+0.7}_{-1.2}$ & $14.4^{+0.3}_{-0.3}$ \\
 \rule{0pt}{3ex}%
 THM & $79.10\pm0.19$ & $8.0^{+0.3}_{-0.4}$ & $11.8^{+0.4}_{-0.4}$ & $2.4^{+0.7}_{-1.2}$ & $14.4^{+0.3}_{-0.3}$\\
 \rule{0pt}{3ex}%
 TPHM PV=300 FS=4 $\ell\leq2$ & $83.47\pm0.14$ & $7.8^{+0.3}_{-0.3}$ & $12.2^{+0.3}_{-0.4}$ & $2.7^{+0.8}_{-1.1}$ & $14.7^{+0.3}_{-0.3}$ \\
 \rule{0pt}{3ex}%
 TPHM PV=300 FS=4 $\ell\leq3$ & $83.45\pm0.19$ & $8.0^{+0.3}_{-0.3}$ & $12.2^{+0.3}_{-0.4}$ & $2.7^{+0.8}_{-1.2}$ & $14.8^{+0.3}_{-0.3}$ \\
 \rule{0pt}{3ex}%
 TPHM PV=300 FS=4 $\ell\leq4$ & $81.93\pm0.24$ & $8.0^{+0.3}_{-0.4}$ & $12.0^{+0.3}_{-0.4}$ & $2.6^{+0.7}_{-1.1}$ & $14.6^{+0.3}_{-0.3}$ \\
 \rule{0pt}{3ex}%
 TPHM PV=300 FS=4 $\ell\leq5$ & $81.90\pm0.21$ & $8.0^{+0.3}_{-0.4}$ & $12.0^{+0.3}_{-0.4}$ & $2.6^{+0.6}_{-1.1}$ & $14.6^{+0.3}_{-0.3}$ \\
 \rule{0pt}{3ex}%
 TPHM PV=300 FS=2 & $81.88\pm0.23$ & $8.0^{+0.3}_{-0.4}$ & $12.0^{+0.3}_{-0.4}$ & $2.6^{+0.7}_{-1.1}$ & $14.6^{+0.2}_{-0.3}$ \\
 \rule{0pt}{3ex}%
 TPHM PV=22311 FS=3 & $81.86\pm0.30$ & $8.0^{+0.3}_{-0.4}$ & $12.0^{+0.3}_{-0.4}$ & $2.6^{+0.8}_{-1.2}$ & $14.6^{+0.3}_{-0.3}$ \\
\end{tabular}
\end{ruledtabular}
\end{table*}

\begin{table*}[tbph!]
\caption{Source properties for GW190521, listed as median posterior values with error estimates given by the $90\%$ credible intervals. The first three results columns correspond to the results reported in \cite{Abbott:2020tfl,Abbott:2020mjq}, the fourth column summarizes results from \cite{Nitz:2020mga}, and the last three columns are the new results from this paper, taken from our LALInference default runs with the standard versions of the \phXPHM and \phTPHM waveform models (including two choices of mode content for \phTPHM, which yield very similar results)}.
\begin{ruledtabular}
\begin{tabular}{l c c c c c c c}
Waveform Model  & \NRSur  & Pv3HM  & v4PHM & XPHM (NC) & XPHM & TPHM $\ell\leq4$ & TPHM $\ell\leq5$  \\
\hline
Primary BH mass $m_1$ & \mOne &  $90^{+23}_{-16}$ & $99^{+42}_{-19}$ & $129^{+46}_{-37}$ & $97^{+34}_{-21}$ & $109^{+80}_{-22}$ & $107^{+68}_{-20}$ \\
\rule{0pt}{3ex}%
Secondary BH mass $m_2$& \mTwo & $65^{+16}_{-18}$ & $71^{+21}_{-28}$ & $32^{+33}_{-17}$ & $59^{+22}_{-25}$ & $65^{+28}_{-34}$ & $68^{+26}_{-33}$\\
\rule{0pt}{3ex}%
Total BBH mass $M$ & \mTotal & $154^{+25}_{-16}$ & $170^{+36}_{-23}$ & $169^{+23}_{-20}$ & $154^{+35}_{-16}$ & $181^{+44}_{-27}$ & $179^{+39}_{-25}$ \\
\rule{0pt}{3ex}%
Binary chirp mass $\mathcal{M}$ & \mChirp & $65^{+11}_{-7}$ & $71^{+15}_{-10}$ & $55^{+14}_{-16}$ & $64^{+15}_{-10}$ & $71^{+16}_{-11}$ & $72^{+16}_{-11}$\\
\rule{0pt}{3ex}%
Mass-ratio $q=m_2/m_1$ & \massRatio & $0.73^{+0.24}_{-0.29}$ & $0.74^{+0.23}_{-0.42}$ & $0.23^{+0.46}_{-0.14}$ & $0.61^{+0.32}_{-0.36}$ & $0.63^{+0.32}_{-0.46}$ & $0.66^{+0.29}_{-0.46}$ \\

\hline
\rule{0pt}{3ex}%
Primary BH spin $\chi_1$ & \aOne  & $0.65^{+0.32}_{-0.57}$ & $0.80^{+0.18}_{-0.58}$ & $0.84^{+0.12}_{-0.46}$ & $0.67^{+0.30}_{-0.59}$ & $0.86^{+0.12}_{-0.56}$ & $0.84^{+0.14}_{-0.56}$\\
\rule{0pt}{3ex}%
Secondary BH spin $\chi_2$ & \aTwo  & $0.53^{+0.42}_{-0.48}$  & $0.54^{+0.41}_{-0.48}$ & $0.57^{+0.32}_{-0.44}$ & $0.55^{+0.4}_{-0.49}$ & $0.56^{+0.39}_{-0.50}$ & $0.56^{+0.39}_{-0.50}$ \\
\rule{0pt}{3ex}%
Primary BH spin tilt angle $\theta_{LS_1}$ & \thOne  & $80^{+64}_{-49}$  & $81^{+49}_{-45}$ &  $132^{+17}_{-54}$ & $117^{+44}_{-81}$ & $80^{+54}_{-32}$ & $85^{+52}_{-37}$\\
\rule{0pt}{3ex}%
Secondary BH spin tilt angle $\theta_{LS_2}$ & \thTwo  & $88^{+63}_{-58}$  & $93^{+61}_{-60}$ & $84^{+48}_{-44}$ & $82^{+68}_{-57}$ & $97^{+57}_{-64}$ & $97^{+57}_{-63}$\\
\rule{0pt}{3ex}%
Effective inspiral spin parameter $\chi_\mathrm{eff}$ & \chiEff & $0.06^{+0.31}_{-0.39}$ & $0.06^{+0.34}_{-0.35}$ & $-0.46^{+0.55}_{-0.14}$ & $-0.11^{+0.43}_{-0.47}$ & $0.07^{+0.32}_{-0.44}$ & $0.02^{+0.36}_{-0.41}$\\
\rule{0pt}{3ex}%
Effective precession spin parameter $\chi_\mathrm{p}$ & \chiP & $0.60^{+0.33}_{-0.44}$ & $0.74^{+0.21}_{-0.40}$ & $0.57^{+0.19}_{-0.25}$ & $0.49^{+0.34}_{-0.34}$ & $0.78^{+0.17}_{-0.39}$ & $0.76^{+0.19}_{-0.39}$\\

\hline
\rule{0pt}{3ex}%
Remnant BH mass $M_\mathrm{f}~(\Msun)$ & \mFinal & $147^{+23}_{-15}$ & $162^{+35}_{-22}$ & - & $148^{+35}_{-15}$ & $173^{+46}_{-25}$ & $171^{+28}_{-19}$\\
\rule{0pt}{3ex}%
Remnant BH spin $\chi_\mathrm{f}$ & \aFinal & $0.72^{+0.11}_{-0.15}$ & $0.74^{+0.12}_{-0.14}$ & - & $0.63^{+0.17}_{-0.24}$ & $0.75^{+0.13}_{-0.18}$ & $0.72^{+0.13}_{-0.14}$\\
\rule{0pt}{3ex}%
Radiated energy $E_\mathrm{rad}$ & \mTotalMinusMfinal & $7.2^{+2.7}_{-2.2}$ & $7.8^{+2.8}_{-2.3}$ & - & $6.1^{+3.4}_{-3.3}$ & $7.2^{+3.1}_{-3.2}$ & $7.3^{+3.0}_{-3.1}$\\
\hline
\rule{0pt}{3ex}%
Luminosity distance $D_\mathrm{L}$ & $5.3^{+2.4}_{-2.6}$  & $4.6^{+1.6}_{-1.6}$ & $4.0^{+2.0}_{-1.8}$ & $2.9^{+4.1}_{-1.4}$ & $3.5^{+2.4}_{-2.0}$ & $3.5^{+1.9}_{-1.7}$ & $3.4^{+2.0}_{-1.6}$ \\
\rule{0pt}{3ex}%
Source redshift $z$ & $0.82^{+0.28}_{-0.34}$ & $0.73^{+0.20}_{-0.22}$& $0.64^{+0.25}_{-0.26}$ & $0.33^{+0.36}_{-0.15}$ & $0.59^{+0.32}_{-0.3}$ & $0.58^{+0.26}_{-0.25}$ & $0.56^{+0.27}_{-0.23}$ \\
\end{tabular}
\end{ruledtabular}
\label{tab:parameters}
\end{table*}

\subsection{Non-precessing approximants}\label{sec:nonprec}

\begin{figure*}[thpb!]
\includegraphics[width=0.66\columnwidth]{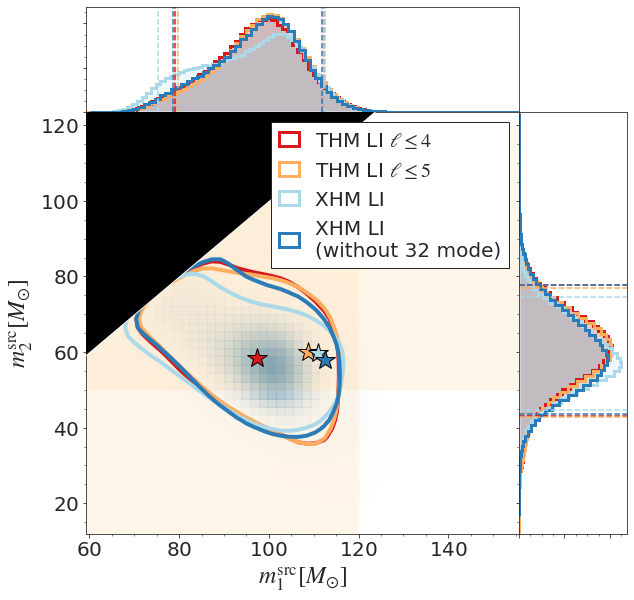}
\includegraphics[width=0.66\columnwidth]{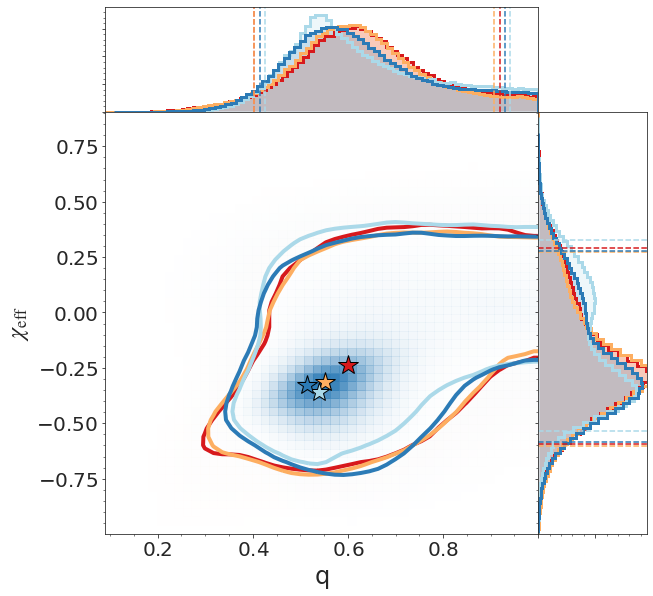}
\includegraphics[width=0.66\columnwidth]{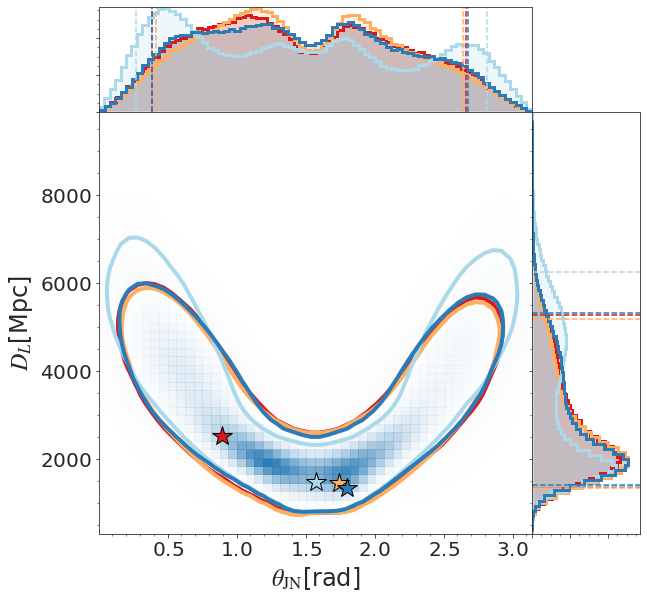}
\caption{Comparison of posteriors with the non-precessing models \phXHM and \phTHM, including two different mode choices for each, obtained with LALInference.}
\label{fig:nonprec-corner}
\end{figure*}

Before turning our attention to precessing models, we will inspect results obtained with non-precessing waveform approximants. In this simplified context, current waveform models have reached a certain level of maturity, where all state-of-the-art versions have been calibrated to NR simulations, including the subdominant harmonics content, to a varying degree. Therefore, we expect good agreement between different models, at least when the same subdominant mode content is included. We show results for LALInference runs with \phTHM and \phXHM in Fig. \ref{fig:nonprec-corner}. One can see that there is consistency between \phXHM and \phTHM when the same set of modes is included, which implies disabling the (3,2) mode in \phXHM and restricting to $\ellmax=4$ in \phTHM. We do observe larger differences when including all the available modes in each model, with a shift towards slightly lower $q$ and mild multimodality in the distance and inclination parameters for \phXHM,
although joint distributions still look broadly consistent with \phTHM. We also notice that the recovered mass ratio and effective spins are consistent with the values reported in the LVC publications. The same goes for other key parameters, such as source-frame masses, distance and inclination (see Fig. 1 and 6 in \cite{Abbott:2020mjq}). For all these results, both component masses lie confidently within the PISN mass gap (at $90\%$ credible intervals).

\subsection{Analysis with \phXPHM}\label{sec:results_xphm}

The results reported in \cite{Nitz:2020mga} for the \phXPHM model were obtained with the {\em default version} of the model (corresponding to MSA Euler angles and final spin version FS=3). The posteriors obtained by \cite{Nitz:2020mga} have non-zero support in regions of parameter space where the direction of the total angular momentum ${\mathbf J}$ flips (see Sec.~\ref{sec:model_descriptions}) and would thus require careful cross-checks for robustness, as discussed previously. This is due to the fact that for the default version we had initially implemented a different behavior as for other options: instead of attempting to track the direction of the total angular momentum ${\mathbf J}$, a warning message was to be printed, alerting the user that the model is less reliable in case of flipped ${\mathbf J}$. After the publication of \cite{Nitz:2020mga} we however realized that the warning messages had not been printed correctly when the calculation of subdominant harmonics was activated. To avoid confusion, we have more recently implemented a change harmonizing the behavior of the different final-spin versions, and the code now always tracks the direction of ${\mathbf J}$ for all parameter settings; this is now also described in the recently updated Sec. IV D of \cite{Pratten:2020ceb}. 

\begin{figure*}[htpb!]
\includegraphics[width=\columnwidth]{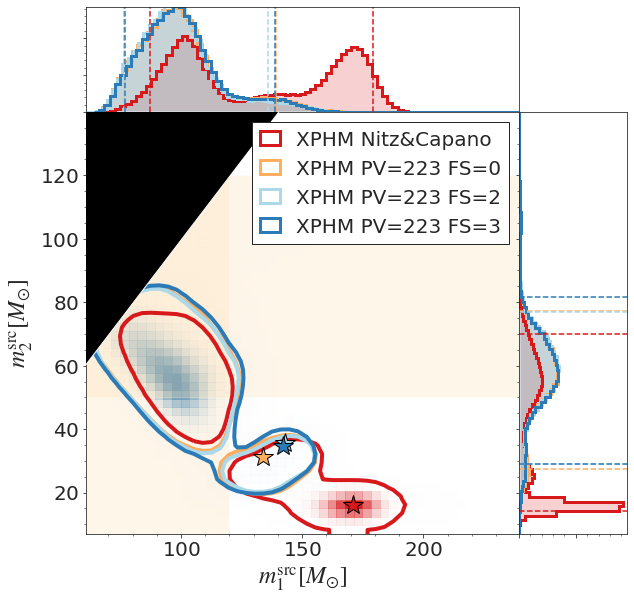}
\includegraphics[width=\columnwidth]{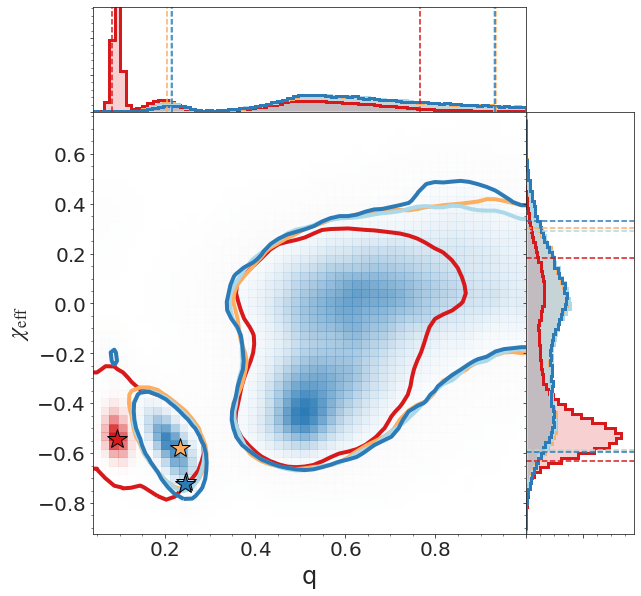}
\includegraphics[width=0.66\columnwidth]{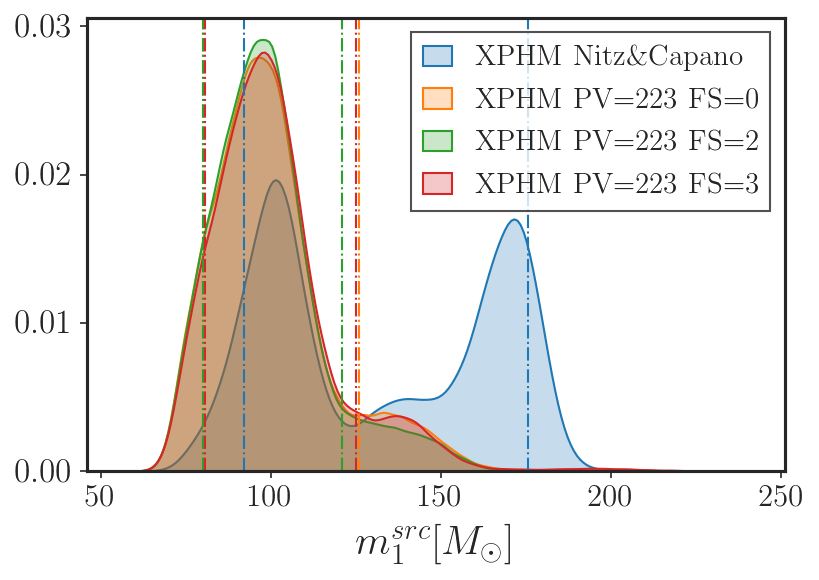}
\includegraphics[width=0.66\columnwidth]{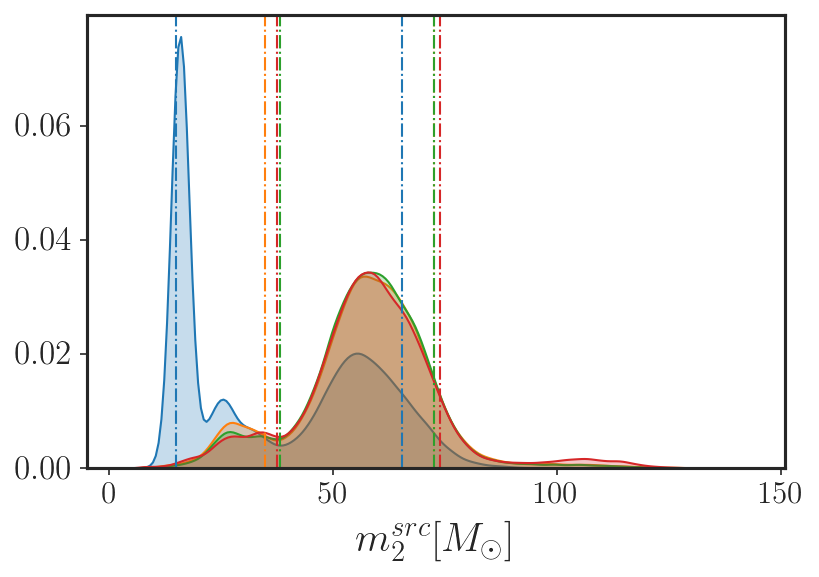}
\includegraphics[width=0.66\columnwidth]{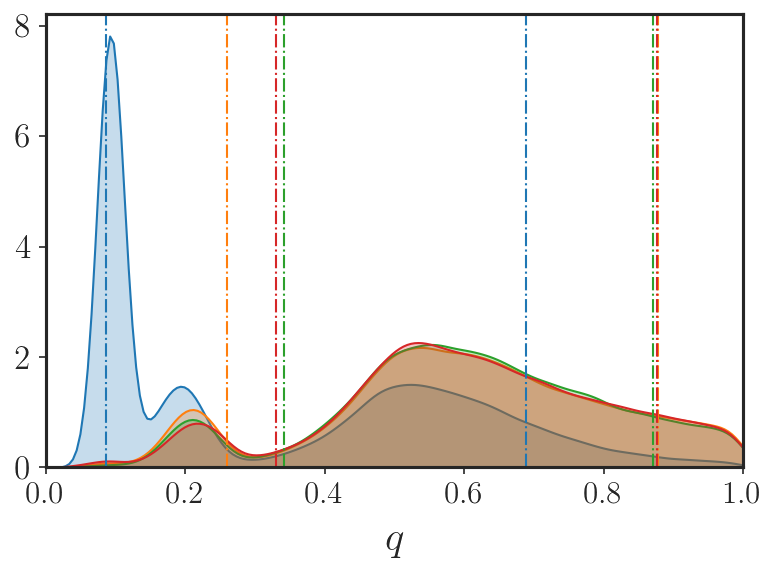}
\caption{Comparison of posterior distributions inferred with pBilby
for different final spin versions of \phXPHM. The prior is uniform in inverse mass ratio $Q$, re-weighted to flat in component masses.
This is however an example where one can see that the $\maxL$ estimate can be less robust than the posterior distribution, with the $\maxL$ point when using the \phPvtwo prescription (FS=0, marked by an orange star) having very different parameters from the $\maxL$ samples obtained using the other alternative final-spin prescriptions (light and dark blue stars) or the one from the Nitz--Capano run (red star). See Sec.~\ref{sec:maxL} for caveats on interpreting $\maxL$ estimates from Bayesian parameter estimation runs. For ease of inspecting the 1D posteriors, we repeat them as additional panels below the corner plot with extended aspect ratio.
}
\label{fig:xphmPB4096}
\end{figure*}

With this change all final-spin versions now produce consistent results, as shown in Fig. \ref{fig:xphmPB4096}, with a much reduced support for the parameter region where the mass ratio is high and the effective spin negative, and where thus ${\mathbf J}$ may flip its sign.
In particular, we note that, using the latest code version, the support for both masses being outside the mass-gap is drastically reduced, see Table~\ref{tab:PISNprob}.
Consistent results with the updated default version have also been reported by~\cite{Capano:2021etf} and \cite{mehta2021observing}, but here we present the first direct comparison using multiple final-spin versions.
We also find in Fig. \ref{fig:xphmPB4096} that when changing the final-spin version, the position of maximum-likelihood sample changes considerably, this is however not surprising as discussed in Sec. \ref{sec:maxL}.

\subsection{Multi-modality and support for high \texorpdfstring{$Q$}{Q}}\label{sec:results_multimode}

We now turn to examining the results obtained with the default settings of \phXPHM and \phTPHM with LALInference and pBilby, where in both cases the two approximants were run with the same priors and sampler settings. As we have already mentioned, pBilby posteriors have been re-weighted to allow a direct comparison with LALInference results, see Sec.~\ref{sec:priors}. Results are shown in Fig.~\ref{fig:PBLIcomparison_mq}. One can appreciate a remarkable consistency between the two sampling codes. It is also clear that mass-ratio posteriors have a multi-modal behavior for both models. The main difference here is that more unequal mass ratios ($q\sim 0.25$) in \phXPHM are correlated with large negative $\chieff$ while the unequal-mass-ratio support for \phTPHM is correlated with moderate positive $\chieff$. Compared to inference with aligned-spin models, support for the components to lie within the mass-gap is reduced (see left panel), however we defer an extensive discussion of this point to Sec.~\ref{sec:results_gap}. In line with~\cite{Nitz:2020mga}, we find evidence for at least one high mass-ratio mode at $Q\approx5$, in addition to the mode with near-equal masses as originally reported by the LVC analysis (right panel).

\begin{figure}[htpb!]
\includegraphics[width=\columnwidth]{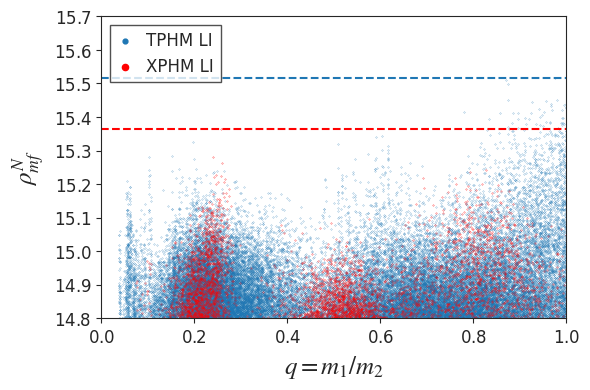}
\caption{Comparison of network matched-filter SNR as a function of mass ratio $q=m_1/m_2$, for our default \phTPHM and \phXPHM results obtained with LALINference MCMC. The dashed lines indicate the maximum SNR values. Only values greater than $14.8$ are shown.}
\label{fig:tphm-xphm-LI-snr}
\end{figure}

In Fig.~\ref{fig:tphm-xphm-LI-snr} we can see a comparison of the highest SNR values for the default LALInference runs with both \phXPHM and \phTPHM. We can observe that the $q\sim 0.2$ region produces similar SNRs for both models, but \phTPHM has support for higher SNRs in the close-to-equal mass region. \phTPHM also recovers a small strip at $q\sim 0.1$ at more or less the same height as the $q\sim 0.2$ bulk. Differences in network matched-filter SNR here are only about 0.15 (which corresponds to 3.86 in $\maxL$). For \phXPHM, the maximum SNR is located at $q=0.26$ while for \phTPHM it is located at $q=0.975$, in agreement with the $\maxL$ positions shown in the right panel of Fig.~\ref{fig:PBLIcomparison_mq}.

\begin{figure}[htpb!]
\includegraphics[width=\columnwidth]{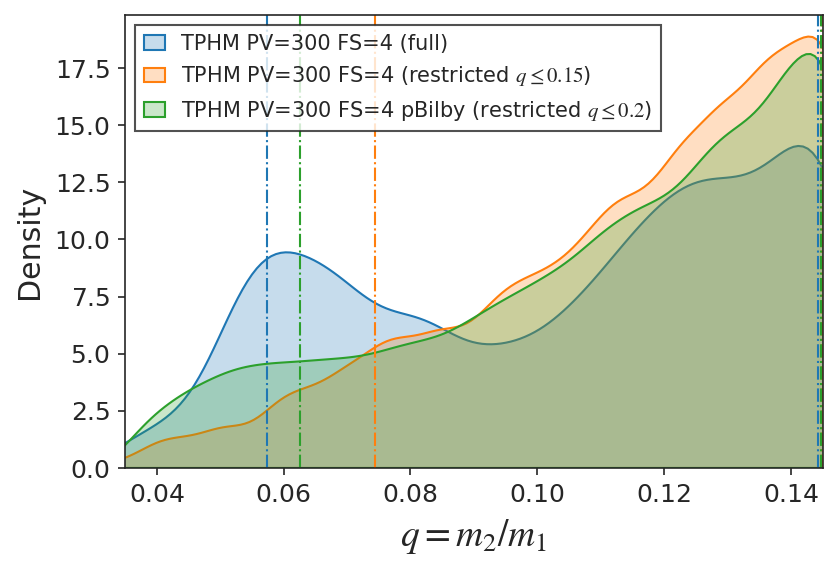}
\includegraphics[width=\columnwidth]{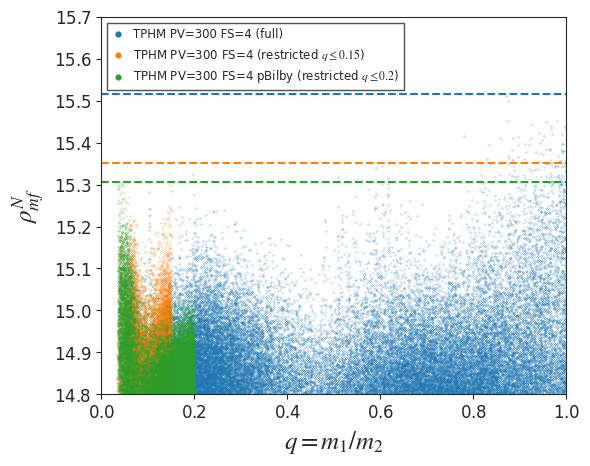}
\caption{Top: Posterior distributions for the mass ratio, comparing the standard \phTPHM results with the full prior range (blue) against restricted-prior results obtained with LALInference (orange) and pBilby (green). Bottom: SNR values as a function of mass ratio for the same three full-prior and restricted-prior runs. Dashed lines correspond to the maximum SNR value from each run. Only values greater than $14.5$ are shown.} 
\label{fig:tphm-restricted}
\end{figure}

In order to better explore the regime of very unequal masses, we have performed \phTPHM runs with restricted mass-ratio priors, both with LALInference and with pBilby. In Fig. \ref{fig:tphm-restricted} one can see that results are consistent with not finding particular support for another mode below the one at $q\sim0.2$. The full prior run is poorly sampled in this region, with only $3.6\%$ of samples below $q=0.15$, so the small peak at $q=0.06$ is probably an artefact from insufficient resolution in this region, and it is not recovered by the restricted runs. The maximum SNR recovered by the restricted runs is also lower than the SNR recovered by the full run near equal masses.

We have also checked that results are robust for different \phTPHM versions, as can be seen in Fig.~\ref{fig:tphm-versions}.
As discussed in appendix~\ref{sec:appendixhm}, the bimodality is not recovered when using only the dominant $\ell\leq2$ spherical harmonic modes, but is robust under inclusion of different subsets of the higher-order modes implemented in \phTPHM.
Therefore, we overall find clear evidence for a bimodal mass-ratio posterior with a secondary unequal mass-ratio peak near $Q\sim5$, but while there is also non-vanishing posterior support reaching to even more extreme ratios, we find no clear evidence for a third peak as originally reported in \cite{Nitz:2020mga}.
These results are consistent with~\cite{Capano:2021etf} and \cite{mehta2021observing}.

\begin{figure*}[htpb!]
\includegraphics[width=0.66\columnwidth]{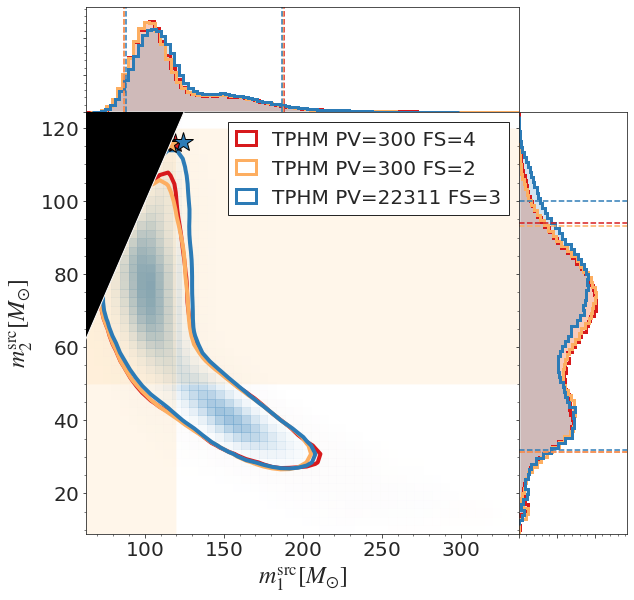}
\includegraphics[width=0.66\columnwidth]{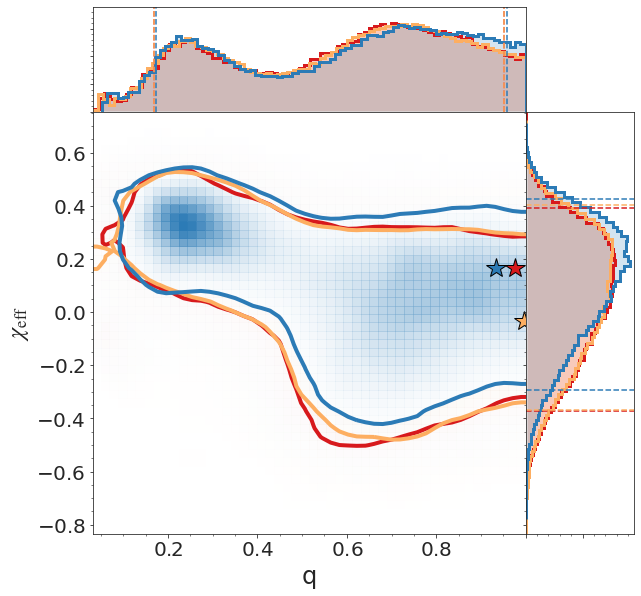}
\includegraphics[width=0.66\columnwidth]{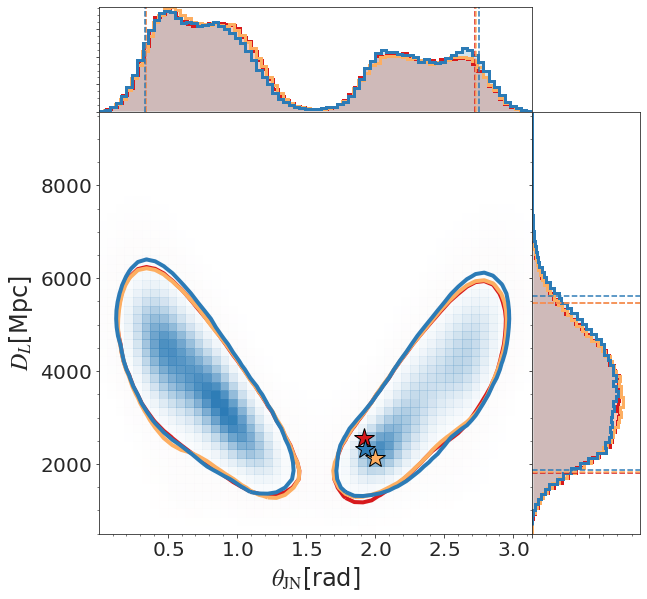}
\caption{Comparison of LALInference results with different precession versions of the \phTPHM model.
\label{fig:tphm-versions}
}
\end{figure*}

\subsection{Spin and precession} 
\label{sec:results_prec}

In terms of spins, one can see in Fig. \ref{fig:PBLIcomparison_spinLoc} that while \phXPHM recovers a posterior that is approximately symmetric around $\chip=0.5$, \phTPHM strongly favors high values of $\chip$. For $\chieff$, the \phXPHM posterior is bimodal, with peaks close to $\chieff=0$ and at large negative values, while the \phTPHM posterior is broad but unimodal with the median close to small positive values. This is reflected also in the position of the $\maxL$ points from the posteriors: while for \phTPHM results, both with pBilby and LALInference, $\maxL$ lies at low positive $\chieff$ and high $\chip \geq 0.8$, for \phXPHM the $\maxL$ is located around $\chip\sim0.5$ and its $\chieff$ is large and negative.

Another thing to notice in relation with Fig.~\ref{fig:PBLIcomparison_spinLoc} is that adding precession does not seem to significantly affect the distance--$\theta_{JN}$ joint distribution for \phXPHM with respect to \phXHM, while for the time-domain \phT model family precession indeed helps to better constrain the posteriors. This is supported by the Bayes factors for the precessing hypothesis (computed from the difference in log signal-to-noise Bayes factors between precessing and non-precessing results) shown in Table~\ref{tab:tabBFprec}, where the time-domain models show stronger support for precession than the frequency-domain models.

\begin{table*}
\begin{ruledtabular}
\begin{tabular}{ccccccc}
 & TPHM vs THM ($\ellmax=4$) & TPHM vs THM ($\ellmax=5$) & XPHM vs XHM & TPHM vs XHM & NRSur7dq4 (LVC)  \\
\hline \hline 
Bayes factor & \ensuremath{17.0_{-4.6}^{+6.2}} & \ensuremath{13.5_{-3.0}^{+3.9}} & \ensuremath{1.5_{-0.3}^{+0.4}} & \ensuremath{6.5_{-1.6}^{+2.2}} & \ensuremath{11.5_{-1.1}^{+1.1}} \\
\end{tabular}
\end{ruledtabular}
\caption{
\label{tab:tabBFprec}
Comparison of Bayes factors between precessing and non-precessing approximants.}
\end{table*}

\subsection{Masses and mass gap}
\label{sec:results_gap}

In Table~\ref{tab:PISNprob} we show the posterior probabilities for the component objects of the source of GW190521 being inside the PISN gap. The exact boundaries of this gap are not known exactly and depend in a highly non-trivial way on nuclear reaction rates and several aspects of stellar dynamics. For simplicity, we will provide probabilities computed assuming a gap either in the range $[50,120]\,M_{\odot}$ \cite{Woosley:2016hmi} or, following more recent estimates \cite{Woosley:2021xba}, in the range $[70,161]\,M_{\odot}$ (we report results for the latter case in parentheses). Shifting the estimated boundaries towards higher masses increases (decreases) the probability of the heavier (lighter) component lying in the gap; it also drastically decreases the probability of both masses being in the mass gap.  

We observe that \phXPHM with a prior that is flat in component masses (run with LI) has more support for the PISN gap hypothesis, as do the non-precessing models (both \phXHM and \phTHM). One can see that for the \phXPHM LI run, sampled uniform in component masses, the support for at least one component in the mass gap is greater than $90\%$. The alternative prior uniform in $Q=1/q$ enhances the high mass-ratio region of the posterior, which typically makes both components lie outside the gap. In the runs we performed with pBilby and this prior, the support for components in the mass gap drops for \phXPHM. With \phTPHM, the mass gap probabilities are generally lower than for \phXPHM.

Hence, we conclude that inference with non-precessing models, or with \phXPHM and the uniform-in-$q$ prior, prefer the hypothesis of at least one object in the PISN gap ($>9:1$ probability ratio), matching the original conclusions of~\cite{Abbott:2020tfl,Abbott:2020mjq}; but \phXPHM runs with uniform-in-$Q$ prior and \phTPHM inference have only mild preference (ranging from $\sim 2:1$ to $\sim 3:1$ depending on model version and mode content) for this scenario, allowing more readily for the ``straddling" hypothesis from \cite{Fishbach:2020qag} of the heavier BH above and the lighter BH below the PISN gap.
As seen in Table \ref{tab:PISNprob}, results for \phTPHM with different mode content ($\ell \leq 3,4,5$) are broadly consistent,
except for a larger probability with $\ell \leq 3$ of having the smaller mass, or either of the two masses, in the mass gap.

\begin{table*}
\begin{ruledtabular}
\begin{tabular}{lcccc}
 & $P(m_{1,\mathrm{src}}^{\mathrm{PISN}})$ & $P(m_{2,\mathrm{src}}^{\mathrm{PISN}})$ & $P(m_{1,\mathrm{src}}^{\mathrm{PISN}} \& m_{2,\mathrm{src}}^{\mathrm{PISN}}) $ & $P(m_{1,\mathrm{src}}^{\mathrm{PISN}} || m_{2,\mathrm{src}}^{\mathrm{PISN}}) $  \\
\hline \hline 
TPHM $\texttt{PV}=300\ \texttt{FS}=4$ $\ell\leq3$ & 0.672(0.919) & 0.843(0.612) & 0.663(0.612) & 0.852(0.919)  \\
\textbf{TPHM $\texttt{PV}=300\ \texttt{FS}=4$ $\ell\leq4$} & 0.666(0.872) & 0.715(0.415) & 0.646(0.415) & 0.736(0.872)  \\
\textbf{TPHM $\texttt{PV}=300\ \texttt{FS}=4$ $\ell\leq5$} & 0.741(0.916) & 0.782(0.460) & 0.719(0.460) & 0.804(0.916)  \\
TPHM $\texttt{PV}=300\ \texttt{FS}=2$ & 0.661(0.871) & 0.705(0.414) & 0.640(0.414) & 0.726(0.871)  \\
TPHM $\texttt{PV}=22311\ \texttt{FS}=3$ & 0.629(0.870) & 0.713(0.463) & 0.613(0.463) & 0.728(0.870)  \\
\textbf{TPHM $\texttt{PV}=300\ \texttt{FS}=4$ fixed sky location} & 0.648(0.807) & 0.644(0.442) & 0.625(0.442) & 0.667(0.808)  \\
\textbf{TPHM $\texttt{PV}=300\ \texttt{FS}=4$ fixed 3D localization} & 0.896(0.914) & 0.897(0.780) & 0.886(0.780) & 0.907(0.914)   \\
THM  & 0.993(0.997) & 0.857(0.182) & 0.856(0.182) & 0.996(0.997)  \\
\hline
\textbf{XPHM N$\&$C} & 0.470(0.675) & 0.372(0.048) & 0.365(0.048) & 0.477(0.675)  \\
\textbf{XPHM LI $\texttt{PV}=223\ \texttt{FS}=3$} & 0.910(0.993) & 0.785(0.166) & 0.750(0.166) & 0.946(0.993)  \\
XPHM PB 1/q $\texttt{PV}=223\ \texttt{FS}=3$ & 0.681(0.923) & 0.495(0.072) & 0.474(0.072) & 0.701(0.924)   \\
XPHM PB 1/q $\texttt{PV}=223\ \texttt{FS}=2$ & 0.711(0.970) & 0.510(0.069) & 0.493(0.069) & 0.728(0.970)   \\
XPHM PB 1/q $\texttt{PV}=223\ \texttt{FS}=0$ & 0.658(0.959) & 0.479(0.067) & 0.464(0.067) & 0.674(0.959)  \\
XHM LI & 0.994(0.993) & 0.847(0.125) & 0.845(0.125) & 0.997(0.993)\\
\end{tabular}
\end{ruledtabular}
\caption{
\label{tab:PISNprob}
Probabilities of component objects to be in the PISN mass gap, assumed to be either in the range $[50,120]\,M_{\odot}$ or in $[70,161]\,M_{\odot}$ (in parentheses). Numbers reported here are computed as the ratio between the posterior samples inside the gap and the total sample size. The runs denoted with $1/q$ use priors uniform in $Q=1/q=m_1/m_2$, while all others use the default prior uniform in $q=m_2/m_1$; see Sec.~\ref{sec:results_gap} for the full discussion.
}
\end{table*}

\subsection{Extrinsic parameters and EM counterpart}
\label{sec:results_EM}

The tentative association of GW190521 with the AGN flare \texttt{ZTF19abanrhr} \cite{Graham:2020gwr} would be hugely impactful as the first detected electromagnetic counterpart to a BBH merger and open up cosmological applications \cite{Chen:2020gek,Mukherjee:2020kki,Gayathri:2020fbl,Finke:2021aom}.
This association is however far from certain \cite{Ashton:2020kyr,Palmese:2021wcv}. Still, below we report updated association probabilities based on our new parameter estimation results. It is also illustrative to study how the GW190521 posteriors would change when factoring in knowledge of the sky location and distance of this counterpart as priors on the GW parameter estimation. To do so, we consider two scenarios: in the first, we only fix the right ascension and declination of the source, while in the second we also fix its luminosity distance, which can be calculated from the flare's redshift assuming a standard cosmology \cite{Planck2015}. 

First, looking back at the $d_L$--$\theta_{JN}$ panel of Fig.~\ref{fig:PBLIcomparison_spinLoc} for the runs with standard priors (no constraints on source location), we can see that while none of the models is able to break the hemisphere degeneracy, \phTPHM results are able to constrain more a particular inclination in each hemisphere, albeit some bimodality in each hemisphere is also present. Similarly, \phTPHM appears to deliver a better constraint on the luminosity distance than \phXPHM, for which there are also hints of bimodality in the posterior. We note also that the $\maxL$ sample of the \phXPHM posterior has a lower luminosity distance than that of the \phTPHM posterior, and that these results are consistent between LALInference and pBilby runs.

\begin{figure*}[htpb!]
\includegraphics[width=\columnwidth]{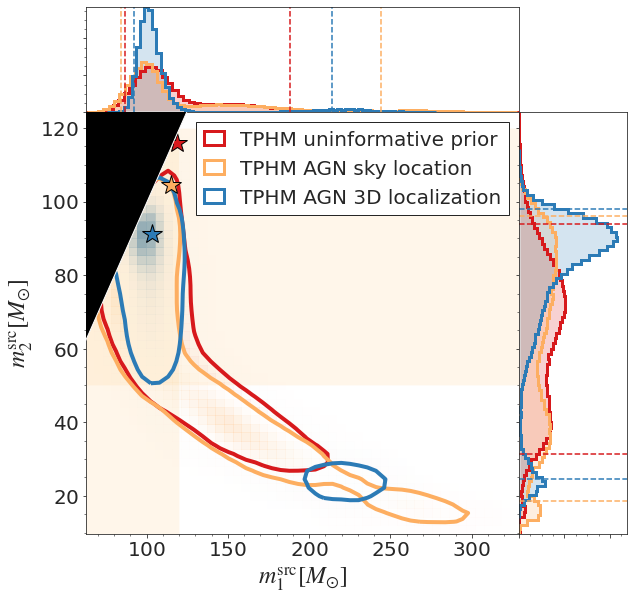}
\includegraphics[width=\columnwidth]{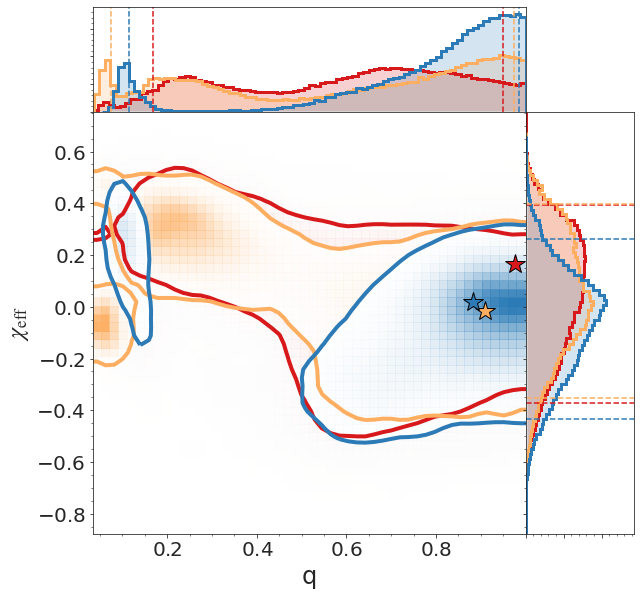}
\caption{Results for the standard runs with uninformative sky localization and distance priors compared with runs where either the 2D sky location or the full 3D localization are fixed to those of the tentative AGN counterpart, for the \phTPHM default version with LALInference.}
\label{fig:tphm-skyloc}
\end{figure*}

Switching to \phTPHM runs with counterpart-informed priors, in Fig.~\ref{fig:tphm-skyloc} we present the results for masses and spins.
Fixing either the 2D sky location or the full 3D localization to those of the AGN flare, the support at mass ratio $Q$ greater than 5 is enhanced in both cases.
However, fixing the 3D localization also constrains the mass posteriors more overall, while fixing only the 2D sky location mostly extends support to higher $Q$ (lower $q$).
The 2D-fixed prior produces a slightly bimodal posterior in $\chieff$ that is however broadly consistent with that from the standard prior, while the run with fixed 3D localization shifts the posterior mode of $\chieff$ towards $0$.
For the precessing spin parameter $\chip$, fixing only the sky position has no noticeable effect, while fixing the 3D localization shifts the recovered posterior distribution to milder values.
It is also worth noticing that fixing the 3D localization increases the probability for the companions to be inside the PISN mass gap (see Table.~\ref{tab:PISNprob}) with respect to the standard run, while fixing only the sky location prior does not seem to have an effect on this.
We also report the Bayes factors for these fixed-prior runs over the default runs in Table~\ref{tab:tabBFskyloc}, which are quite high, but require a full analysis of the actual association probability to interpret.

\begin{figure}[htpb!]
\includegraphics[width=\columnwidth]{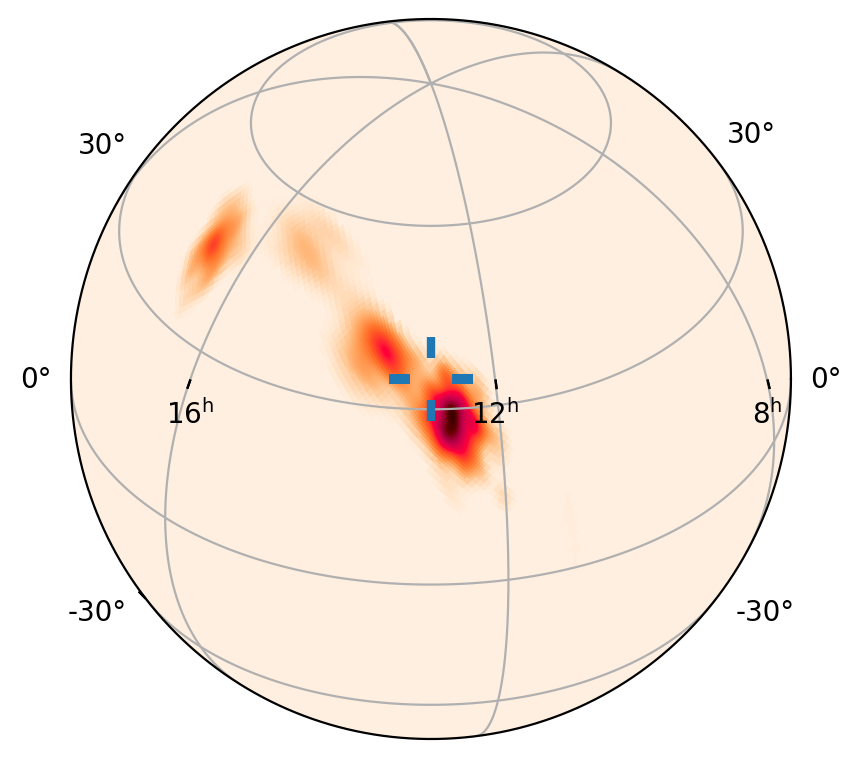}
\caption{Position of the AGN flare ZTF19abanrhr \cite{Graham:2020gwr} compared with the sky location posterior density for GW190521 recovered by our default \phTPHM run.}
\label{fig:skymap}
\end{figure}

To quantify this probability of a physical association between the GW and EM signals, in Table~\ref{tab:oddsEM} we show the results from performing the same multi-messenger coincidence significance analysis as presented in \cite{Ashton:2020kyr} (and using their public code), based on the localization overlap. We give results for the posterior overlap integrals for the sky location ($\mathcal{I}_{\Omega}$), for the distance alone ($\mathcal{I}_{D_{L}}$), and for the combined 3D localization ($\mathcal{I}_{D_{L}}\mathcal{I}_{\Omega}$). Assuming the same prior odds as in \cite{Ashton:2020kyr}, based on the reported number of flares similar to ZTF19abanrhr in the ZTF alert stream, we compute the odds $\mathcal{O}_\mathrm{C/R}$ for the coincident hypothesis.
The resulting odds vary by a factor $\sim2$ between runs with \phXPHM and different versions of \phTPHM (precession prescriptions, final spin versions and mode content) and overall show mild preference for association.
However, our results are consistent with the findings of \cite{Ashton:2020kyr}, with the highest $\mathcal{O}_\mathrm{C/R}$ we find at the same level as that reported for a run with the \seobnrvforphm model in \cite{Ashton:2020kyr}:
essentially, despite the high Bayes factors the evidence is insufficient to confidently associate the events.
For further illustration, Fig.~\ref{fig:skymap} shows the position of ZTF19abanrhr compared with the sky location posterior density for GW190521 recovered by our default \phTPHM run.

\begin{table}
\begin{ruledtabular}
\begin{tabular}{ccc}
 & BF fixed sky location & BF fixed 3D location  \\
\hline \hline 
TPHM  $\ellmax=4$   & \ensuremath{14_{-5}^{+7}}   & \ensuremath{62_{-14}^{+18}} \\
TPHM  $\ellmax=5$   & \ensuremath{13_{-5}^{+7}}   & \ensuremath{64_{-16}^{+21}} \\
XPHM $\ellmax=4$    & \ensuremath{23_{-3}^{+3}}   & \ensuremath{95_{-13}^{+15}} \\
\end{tabular}
\end{ruledtabular}
\caption{
\label{tab:tabBFskyloc}
Comparison of Bayes factors for the runs with either 2D sky location or full 3D localization (sky position plus luminosity distance) fixed to the AGN counterpart candidate~\cite{Graham:2020gwr}, against the default runs with unconstrained localization priors.
While these are quite high, it is important to take into account a full analysis of the multi-messenger coincidence significance, see the text and Table~\ref{tab:oddsEM}.
}
\end{table}

\begin{table}
\begin{ruledtabular}
\begin{tabular}{rcccc}
 &  $\mathcal{I}_{D_{L}}$ & $\mathcal{I}_{\Omega}$ & $\mathcal{I}_{D_{L}}\mathcal{I}_{\Omega}$ & $\mathcal{O}_\mathrm{C/R}$ \\
\hline \hline 
TPHM $\texttt{PV}=300\ \texttt{FS}=4$ $\ellmax=4$ & 4.7 & 17 & 140 & 6 \\
TPHM $\texttt{PV}=300\ \texttt{FS}=4$ $\ellmax=5$ & 5.1 & 30 & 140 & 12 \\
TPHM $\texttt{PV}=300\ \texttt{FS}=2$ $\ellmax=4$ & 4.6 & 33 & 140 & 12\\
TPHM $\texttt{PV}=22311\ \texttt{FS}=3$ $\ellmax=4$ & 4.7 & 20 & 110 & 7 \\
XPHM LI $\texttt{PV}=223\ \texttt{FS}=3$ & 3.8 & 19 & 110 & 6\\
\end{tabular}
\end{ruledtabular}
\caption{
\label{tab:oddsEM}
Posterior overlap integrals and odds for the association between GW190521 and ZTF19abanrhr~\cite{Graham:2020gwr}, following the method from~\cite{Ashton:2020kyr}.
}
\end{table}

\section{Conclusions}
\label{sec:conclusions}

In this paper we have re-analyzed GW190521, the highest-mass GW event yet detected, with two recently developed waveform models: First \phXPHM, which is a successor to previous frequency-domain \phIMR models, which have become standard tools in GW data analysis, but are not an optimal choice for very high-mass events, where SNR primarily comes from the ringdown to the final Kerr black hole, as can be seen in Fig. \ref{fig:waveforms}. Second, we have used the new time-domain \phTPHM model, which improves over \phXPHM in how it treats precession, in particular regarding the ringdown, and which recovers a higher SNR (however consistent within statistical errors) and signal-to-noise Bayes factor (see Table~\ref{tab:SNR_BF_table}).
Our overall results are broadly consistent with the original LVC analysis \cite{Abbott:2020tfl,Abbott:2020mjq}, in the sense that the inferred source parameters are consistent at 90\% credible intervals. We confirm the complicated multi-modal posterior structure as first reported by \cite{Nitz:2020mga}, including support for more unequal mass ratios,  but with different statistical significance of the peaks compared to what has been found in \cite{Nitz:2020mga}.

Before summarizing some more of our findings, we stress that our analysis is not the final answer on this uniquely challenging event either: not only are future waveform models expected to improve accuracy in this regime, but we also expect significant progress in understanding the systematic errors of precessing waveform models. One way of analyzing systematic errors would be to perform injection studies, where NR waveforms are injected into synthetic noise and then the bias of the recovered parameters can be studied. For GW190521-like signals, one of the challenges of such a study is the lack (or sparsity) of suitable NR waveforms across the relevant part of parameter space, including very unequal masses. Another one is that, especially for signals with strong precession, the results may significantly depend on the extrinsic parameters chosen for the injection, which will increase the computational cost and difficulty of interpretation for such studies. 
 
As a first step, it will be interesting to repeat our analysis with other currently available time-domain waveform models which also cover large mass ratios $Q$ and can therefore be used to test posterior support in that region. A forthcoming study \cite{AjitSEOB} will present a similar re-analysis of GW190521 with the \seobnrvforphm model, which will allow a direct comparison with our results.

Meanwhile, the consistency we observe between two independent sampling codes, pBilby using nested sampling \cite{Ashton:2018jfp,Smith:2019ucc} and LALInference \cite{Veitch:2014wba} using MCMC sampling, gives us confidence in our results, which also appear to be robust under changes of mass priors.
We have also exploited the flexibility of the \phXPHM and \phTPHM models to check the influence of different modelling prescriptions and choices of spherical harmonic mode content on parameter estimation, finding that posteriors are generally robust against changes in the models' internal options. In particular, we focused on the final spin approximation, which is crucial for the ringdown regime. We have also discussed caveats when considering $\maxL$ points in parameter space, pointing out that maximum likelihood is typically poorly resolved by the posterior samples from Bayesian parameter estimation algorithms.

While keeping in mind that future waveform models will provide further insight into the properties of GW190521, with this study we already provide improved parameter estimation results for the source of this event.
One of our central results is to confirm the multi-modal nature of the posterior found in \cite{Nitz:2020mga}, with strong support for two peaks at near-equal mass ratios and at $Q\sim5$.
We do however not find significant support for an intermediate mass-ratio merger with mass ratios of ten or higher, consistent with other recent results by~\cite{Capano:2021etf} and \cite{mehta2021observing}.
We do not exclude further modes at higher $Q$, but we do not expect them to be significant based on our findings, and most importantly current waveform models are at this time significantly less reliable for mass ratios below our cutoff of $q=0.035$ ($Q=28.57$).
We also provide updated probabilities of component masses being located in the PISN mass gap, in general confirming the preference of the original LVC analyses for at least one component inside the gap, but with \phTPHM results reducing the preference against a ``straddling" binary configuration with one component below and one above the mass gap; and of associating GW190521 to the \cite{Graham:2020gwr} potential electromagnetic counterpart.

\section*{Acknowledgements}

We thank Gregory Ashton, Colm Talbot and Rory Smith for valuable discussions about GW parameter estimation with Bilby.
This work was supported by European Union FEDER funds,
the Ministry of Science, Innovation and Universities and the Spanish Agencia Estatal de Investigaci{\'o}n grants PID2019- 106416GB-I00/AEI/10.13039/501100011033, 
RED2018- 102661-T, RED2018-102573-E, FPA2017-90687-REDC, 
Comunitat Autonoma de les Illes Balears through the Direcci{\'o}  General
de Pol{\'i}tica Universitaria i Recerca with funds from the Tourist Stay Tax Law ITS 2017-006 (PRD2018/24),
and Conselleria de Fons Europeus, Universitat i Cultura, i Fons Social Europeu,
EU COST Actions CA18108, CA17137, CA16214, and CA16104. 
Generalitat Valenciana (PROMETEO/2019/071).
D. K. is supported by the Spanish Ministerio de Ciencia, Innovaci{\'o}n y Universidades (ref. BEAGAL 18/00148) and cofinanced by the Universitat de les Illes Balears.
M.C. acknowledges funding from the Spanish Agencia Estatal de Investigaci{\'o}n, grant IJC2019-041385.
The authors thankfully acknowledge the computer resources at MareNostrum and the technical support provided by Barcelona Supercomputing Center (BSC) through Grants 
No. 
AECT-2020-2-0015,  
AECT-2020-1-0025,  
AECT-2019-3-0020,  
AECT-2019-2-0010,  
AECT-2019-2-0017,  
AECT-2019-1-0022,  
from the Red Española de Supercomputación (RES).
The authors also acknowledge the computational resources at the cluster CIT provided by LIGO Laboratory and supported by National Science Foundation Grants PHY-0757058 and PHY-0823459.
This research has made use of data obtained from the Gravitational Wave Open Science Center~\cite{GWOSC}, a service of LIGO Laboratory, the LIGO Scientific Collaboration and the Virgo Collaboration. LIGO is funded by the U.S. National Science Foundation. Virgo is funded by the French Centre National de Recherche Scientifique (CNRS), the Italian Istituto Nazionale della Fisica Nucleare (INFN) and the Dutch Nikhef, with contributions by Polish and Hungarian institutes.

\appendix

\section{Analysis of harmonic content}\label{sec:appendixhm}

\begin{figure*}[ptbh!]
\includegraphics[width=.95\columnwidth]{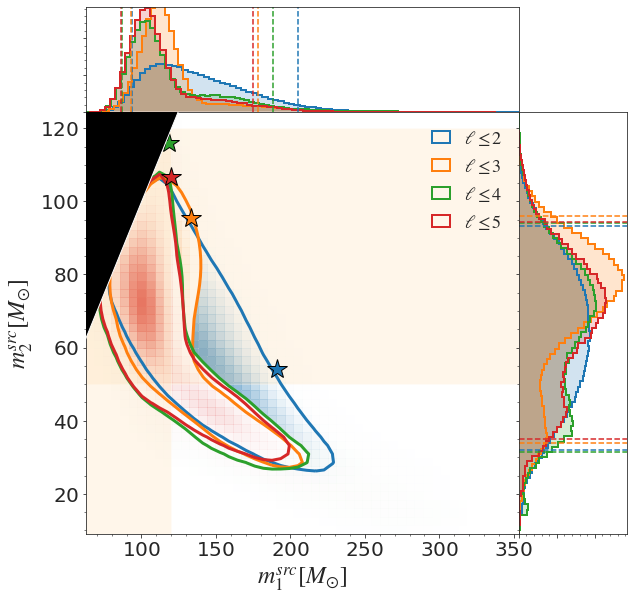}
\includegraphics[width=.95\columnwidth]{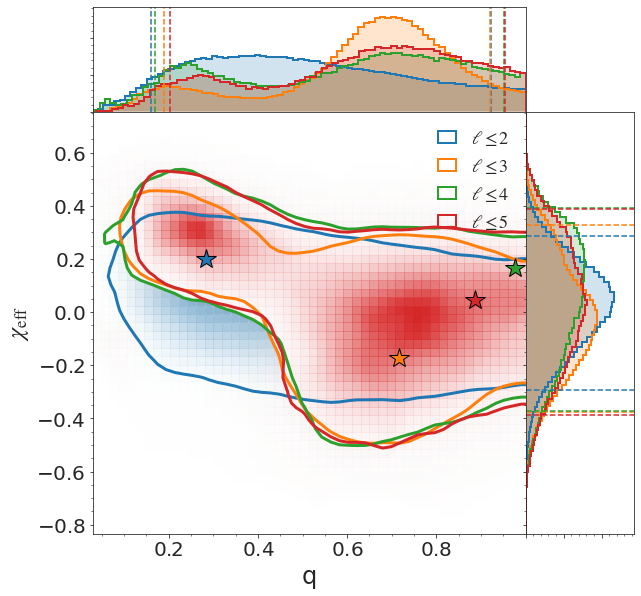}
\includegraphics[width=.95\columnwidth]{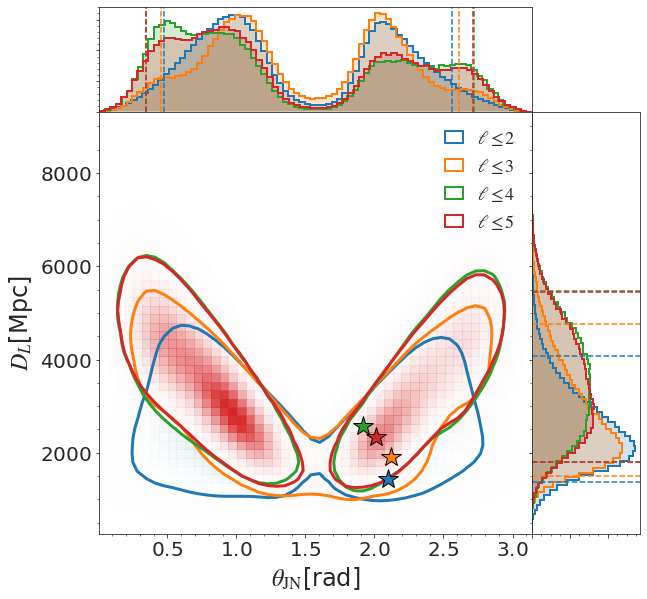}
\includegraphics[width=.95\columnwidth]{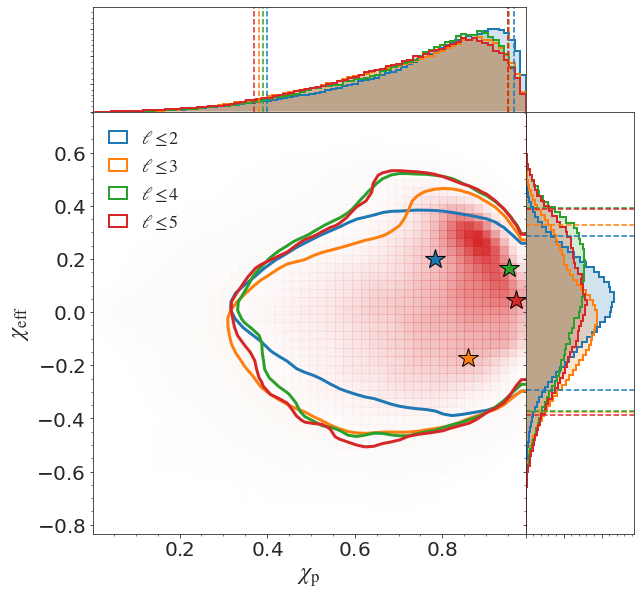}
\caption{
\label{fig:lmaxcomparison}
Comparison of posterior distributions obtained with \phTPHM with different spherical harmonic mode content $\ell\leq2,3,4,5$.}
\end{figure*}

For short-duration signals like GW190521, the inclusion of different harmonics in the signal templates can be even more crucial than in the lower-mass case when the inspiral is also observable, helping to break possible degeneracies between parameters and to obtain more information from the signal. To test the effect of including different sets of harmonics, in Fig.~\ref{fig:lmaxcomparison} we compare the main parameters of the signal for runs performed with different harmonic content, going from $\ell\leq2$ up to $\ell\leq5$. We can observe how the recovered posterior densities tend to converge as more harmonics are included, resulting in a very similar distribution for $\ell\leq4$ and $\ell\leq5$. This finding justifies in part our choice of performing the main runs of this work for the subset $\ell\leq4$. Going beyond $\ell\leq5$, future work will be required to study the impact of harmonics that are not included in our waveform models.

\begin{figure*}[ptbh!]
\includegraphics[width=.95\columnwidth]{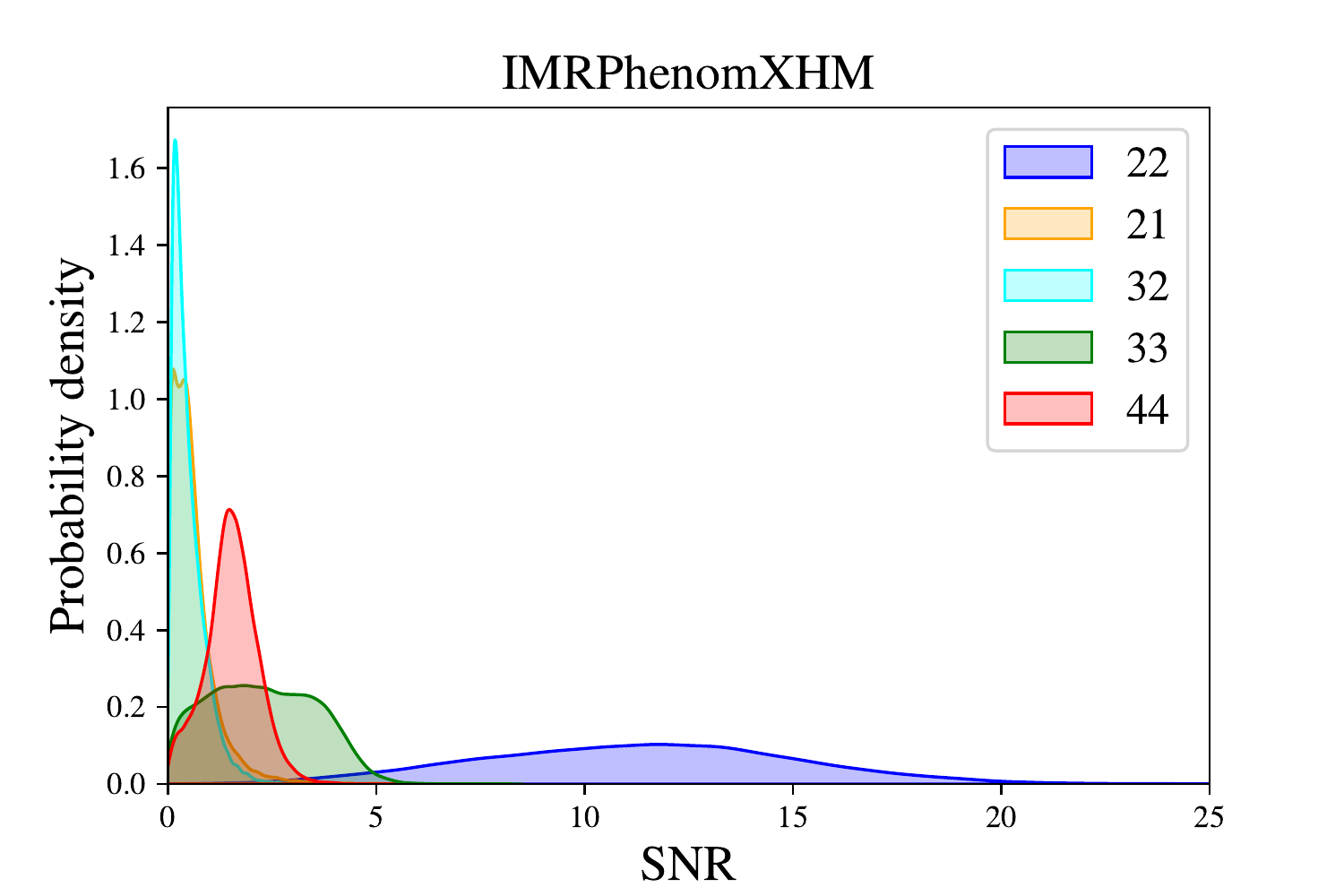}
\includegraphics[width=.95\columnwidth]{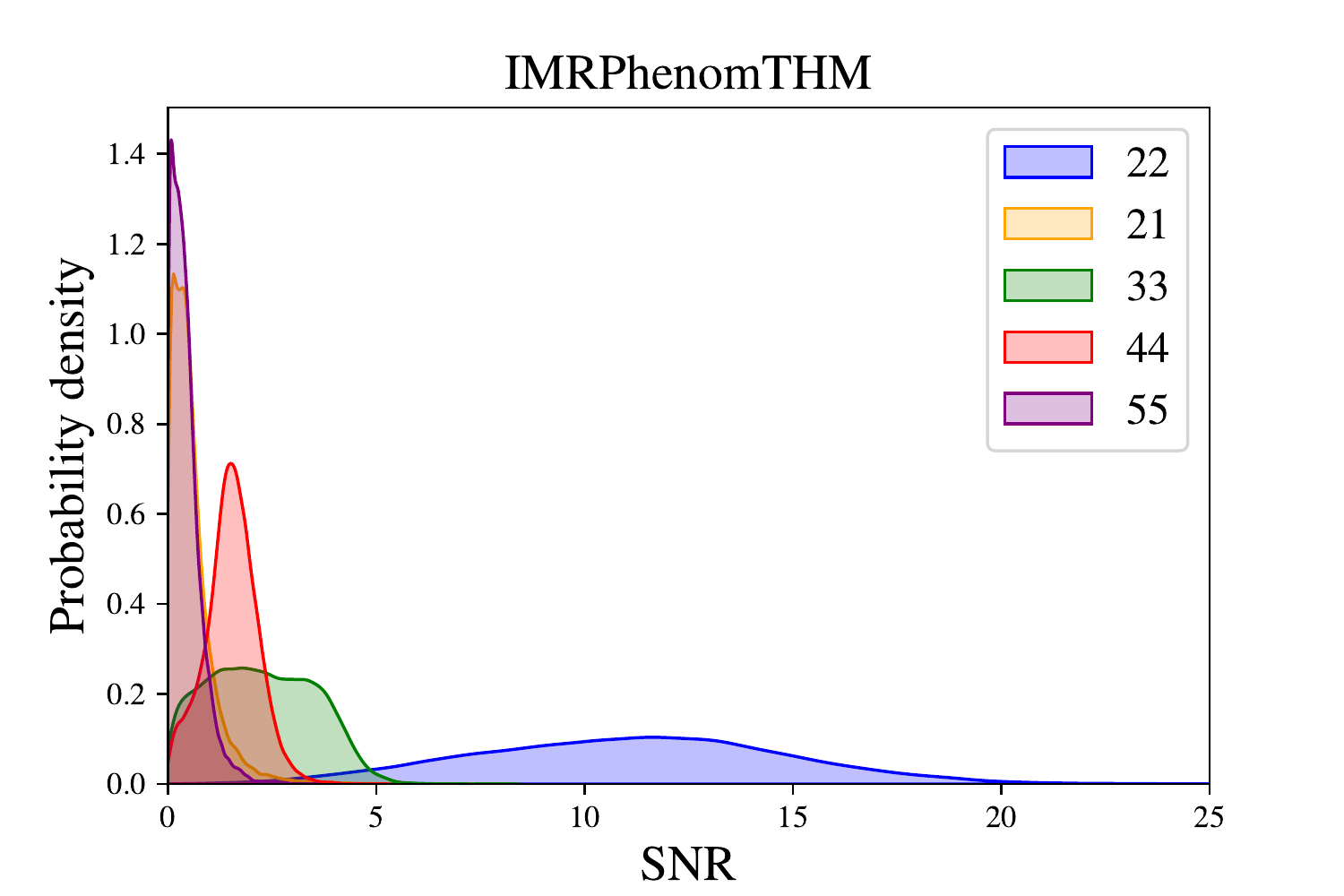}
\caption{Optimal SNR corresponding to different harmonics, computed on the same posterior samples from our default LALInference runs for \phXHM (left panel, with PV=223, FS=3 settings) or \phTHM (right panel, with PV=300,FS=4 settings).
}
\label{fig:mode-snr}
\end{figure*}

A rough quantitative estimate of the importance of higher modes can be obtained by computing the optimal SNR contributed by each harmonic. The optimal SNR for one mode can be defined as 
\begin{equation}
\label{eq:inner_prod}
    \mathrm{\rho^{lm}_\mathrm{opt}} = \sqrt{\Braket{h^{lm} , h^{lm}}},
\end{equation}
where $\Braket{\:,\:}$ refers to the usual noise-weighted inner product and $h^{lm}$ denotes the contribution of a given mode to the strain measured by the detector. We compute the SNR contribution of each harmonic employing the same PSDs we use for parameter estimation throughout this paper. In order to cleanly separate between different modes in the inertial frame, we perform the calculation using the aligned-spin version of the models, namely \phXHM and \phTHM. The source parameters are taken from the posterior samples of the same \phTPHM LALInference run (corresponding to the red curves of Fig.~\ref{fig:PBLIcomparison_spinLoc}). We present our results in Fig.~\ref{fig:mode-snr}: based on the SNR contribution of each harmonic, one would expect the largest impact on posteriors from the inclusion of the (3,3) mode. This conclusion is consistent with the results of a full Bayesian inference analysis outlined above, where the results for $\ell\leq2$ are very different from the more complete runs but changes from $\ell\leq3$ to $\ell\leq4$ are more moderate and the $\ell\leq4$, $\ell\leq5$ results are very similar.

\FloatBarrier


\vspace{0.1in}

\vfil

\let\c\Originalcdefinition %
\let\d\Originalddefinition %
\let\i\Originalidefinition

\bibliography{phenomx,phenom_refs,eob_refs,nr_refs,postnewtonian,gravitationalwaves}


\end{document}